%% file: main_v4.tex
\documentclass[%
 reprint,
 amsmath,amssymb,
 aps,
]{revtex4-2}
\usepackage{graphicx}
\usepackage{dcolumn}
\usepackage{bm}
\usepackage{braket}
\usepackage{color,comment}
\usepackage{amsfonts,amsmath,amsthm,dsfont}
\usepackage[export]{adjustbox}
\usepackage{soul}

\usepackage[linesnumbered,ruled,vlined]{algorithm2e}

\definecolor{some-color}{rgb}{0.75, 0.25, 0.75}

\usepackage{comment}

\begin{document}

\preprint{APS/123-QED}

\title{Dynamical Logical Qubits in the Bacon-Shor Code}
\author{M. Sohaib Alam}
\email{sohaib.alam@nasa.gov}
\affiliation{Quantum Artificial Intelligence Laboratory (QuAIL), NASA Ames Research Center, Moffett Field, CA, 94035, USA}
\affiliation{USRA Research Institute for Advanced Computer Science (RIACS), Mountain View, CA, 94043, USA}

\author{Eleanor Rieffel}
\email{eleanor.rieffel@nasa.gov}
\affiliation{Quantum Artificial Intelligence Laboratory (QuAIL), NASA Ames Research Center, Moffett Field, CA, 94035, USA}
\date{\today}

\begin{abstract}
The Bacon-Shor code is a quantum error correcting subsystem code composed of weight 2 check operators that admits a single logical qubit, and has distance $d$ on a $d \times d$ square lattice. We show that when viewed as a Floquet code, by choosing an appropriate measurement schedule of the check operators, it can additionally host several dynamical logical qubits. Specifically, we identify a period 4 measurement schedule of the check operators that preserves logical information between the instantaneous stabilizer groups. 
Such a schedule measures not only the usual stabilizers of the Bacon-Shor code, but also measures and promotes gauge operators of the parent subsystem code to additional temporary stabilizers that protect the dynamical logical qubits against errors.
We show that the code distance of these Floquet-Bacon-Shor codes scales as $\Theta(d/\sqrt{k})$ on an $n = d \times d$ lattice with $k$ dynamical logical qubits, along with the logical qubit of the parent subsystem code. Unlike the usual Bacon-Shor code, the Floquet-Bacon-Shor code family introduced here can therefore saturate the subsystem bound $kd = O(n)$. Moreover, several errors are shown to be self-corrected purely by the measurement schedule itself. This work provides insights into the design space for dynamical codes and expands the known approaches for constructing Floquet codes.
\end{abstract}

\maketitle

\section{Introduction}
Quantum computers offer a promising avenue for solving computational problems that may be intractable for classical computers. However, in order to be practically useful, one must correct for errors that accrue over the course of the computation. Quantum error correction (QEC) provides a broad framework in which the logical information is encoded in part of the full physical Hilbert space, and can be protected against errors through measurements of appropriate syndrome observables followed by post-measurement correction operations. A leading candidate for QEC has been the surface code \cite{KITAEV20032,bravyi1998quantum,PhysRevA.86.032324}, for its relatively high threshold as well as a simple square lattice architecture.

Recently, a remarkable class of QEC codes was introduced by Hastings and Haah \cite{Hastings2021dynamically} in which the logical degrees of freedom do not form a fixed subspace of the physical Hilbert space, but rather evolve dynamically. The first such example involved the Kitaev honeycomb model \cite{KITAEV20062}, which when viewed as a subsystem code, encodes no logical qubits. However, by choosing an appropriate measurement schedule of the weight-2 check operators, it was shown that this model leads to dynamical logical degrees of freedom. Due to the periodicity of the measurement schedule and the induced instantaneous stabilizer groups (ISGs), such codes are typically referred to as Floquet codes.

The original honeycomb Floquet code \cite{Hastings2021dynamically} was defined on a hexagonal torus, while generalizations to 3-colorable and 3-valent planar graphs were explored in \cite{vuillot2021planar}. Due to a non-trivial automorphism between electric and magnetic operators across measurement cycles, adding boundaries to the model is non-trivial, but were introduced in \cite{Haah2022boundarieshoneycomb}. Benchmarking studies \cite{Gidney2021faulttolerant,Gidney2022benchmarkingplanar,PRXQuantum.4.010310,hilaire2024enhancedfaulttolerancephotonicquantum} showed a competitive threshold for the model, with at least one small scale experiment\cite{wootton2022measurements} demonstrating stabilizer measurements. Since then, several examples of Floquet codes have been constructed, including those without parent subsystem codes\cite{PRXQuantum.4.020341}, those involving the color code\cite{kesselring2022anyon,Townsend_Teague_2023}, constructions based on hyperbolic geometries\cite{higgott2023constructions,fahimniya2023faulttolerant}, in 3D Euclidean space based on fracton order\cite{PhysRevB.108.205116}, involving qudits \cite{tanggara2024simpleconstructionquditfloquet}, using a path integral approach \cite{Bauer_2024,bauer2024xyfloquetcodesimple}, rewinding the measurement schedule\cite{dua2023engineering}, using twists in the geometry\cite{ellison2023floquet}, as well as those tailored for biased noise \cite{setiawan2024tailoringdynamicalcodesbiased}. Generalizations of Floquet code constructions have also been explored using the ZX calculus\cite{bombin2023unifying,Coecke_2011,rodatz2024floquetifyingstabilisercodesdistancepreserving,delafuente2024dynamicalweightreductionpauli}, adiabatic paths\cite{PhysRevB.106.085122}, anyon condensation\cite{kesselring2022anyon}, as well as aperiodic measurement schedules in the form of dynamic automorphism codes\cite{davydova2023quantum}.

A main aim of this work is to provide insights into the design space for dynamical codes. While the dynamics of previously constructed Floquet codes typically requires a description of the embedded toric codes in its ISGs, and may perhaps be intuitively understood through adiabatic paths \cite{PhysRevB.106.085122} or anyon condensation \cite{kesselring2022anyon}, the construction of the Floquet code we describe here follows directly from the subsystem structure of the parent Bacon-Shor code, without the need to move to a condensed matter perspective. In this sense, it provides a complementary approach to the design of dynamical codes and offers a comparatively simpler example of a Floquet code. This code also comes naturally equipped with a boundary, without the complication of non-trivial automorphisms between electric and magnetic operators \cite{Hastings2021dynamically,Haah2022boundarieshoneycomb,vuillot2021planar}. Furthermore, this Floquet code is naturally defined on a square lattice 
which may be appealing from an experimental perspective, and has a period 4 measurement schedule. 
A Floquet code on a square lattice was previously constructed in \cite{Townsend_Teague_2023} by transforming the 2D hexagonal color code \cite{bombin2006topological} to a square lattice geometry using the ZX calculus \cite{coecke2011interacting}, resulting in a period 13 measurement schedule involving 1 and 2 body checks, mapping detectors and logical operators of the original color code to those of the new code. Furthermore, 
square lattice construction may be possible using a path integral approach \cite{davydovasimonstalk,bauer2025x+}, where the circuit acts on a 2D square lattice, alternating between horizontal CX gates and vertical nearest-neighbor $ZZ$ and $XX$ parity measurements in a period-6 schedule, and realizes the behavior of the toric code without directly measuring its stabilizers. In the code described here, the square lattice construction appears more simply and naturally from the subsystem structure of the parent Bacon-Shor code, with the stabilizers and logical operators being readily identifiable.
It is an example of a CSS Floquet code\cite{PRXQuantum.4.020341,kesselring2022anyon}, since all the stabilizers are of either $X$ or $Z$ type. Moreover, unlike the original honeycomb code\cite{Hastings2021dynamically}, it also serves as an example of a Floquet code whose parent subsystem code hosts a non-zero number of logical qubits, 
which can then be augmented by our construction to
host several dynamical logical qubits. Perhaps most importantly, it demonstrates the construction of a Floquet code using essentially only the subsystem structure of the parent code through the addition of gauge defects. We strongly suspect that this approach may be more widely applicable in constructing more examples of Floquet codes.

This paper is organized as follows. In Section II, we recall and describe the usual Bacon-Shor code\cite{PhysRevA.73.012340} using the virtual qubit framework that sets up the notation and makes the discussion in the following sections simpler. In Section III, we show how to introduce logical dynamical degrees of freedom through gauge defects, in addition to the usual Bacon-Shor logical qubit. In Section IV, we discuss error correction for such Floquet-Bacon-Shor codes. We conclude in Section V, with some thoughts on open problems and future work.

\section{Bacon-Shor Code}
\label{secn:bacon-shor}
Here, we briefly review the Bacon-Shor code \cite{PhysRevA.73.012340}. We start by first recalling some basic facts about subspace, or stabilizer, codes as well as subsystem codes. We use the mathematical formalism of `virtual' qubits \cite{PhysRevLett.95.230504,nikobreu-thesis} here since that will be a particularly convenient framework for discussing our construction of Floquet-Bacon-Shor codes. First, note that the Pauli group of $n$ qubits $\mathcal{P}_{n} = \langle i\mathbb{I}, X_1, Z_1, \dots, X_n, Z_n\rangle$ can be transformed under an automorphism $X_i, Z_j \rightarrow \overline{X}_i, \overline{Z}_j$ as $\langle i\mathbb{I}, \overline{X}_1, \overline{Z}_1, \dots, \overline{X}_n, \overline{Z}_n \rangle$ as long as the transformed qubit operators satisfy the canonical commutation relations $[\overline{X}_i, \overline{Z}_j ] = 2 \delta_{ij} \overline{X}_i \overline{Z}_j$ and $\{ \overline{X}_i, \overline{Z}_j \} = 2(1 - \delta_{ij}) \overline{X}_i \overline{Z}_j$. The transformed operators $\{\overline{X}_i, \overline{Z}_i \}_{i=1}^{n}$ are each, in general, products of the physical Pauli operators. These logical operators define `virtual qubits' (as is standard in Heisenberg or stabilizer view), of which there are exactly as many as there are physical qubits, but each virtual qubit is a multiqubit state from the point of view of the physical qubits; the set of virtual qubits represent collective degrees of freedom over the physical qubits. There are myriad ways that one can construct virtual qubits. In the following, we describe an explicit choice of virtual qubits in the Bacon-Shor code and their corresponding logical $\overline{X}$ and $\overline{Z}$ operators that will be convenient for designing Floquet versions of Bacon-Shor codes.

\subsection{Stabilizer and subsystem codes}

\begin{figure*}[!t]
    \centering
\includegraphics[width=\linewidth]{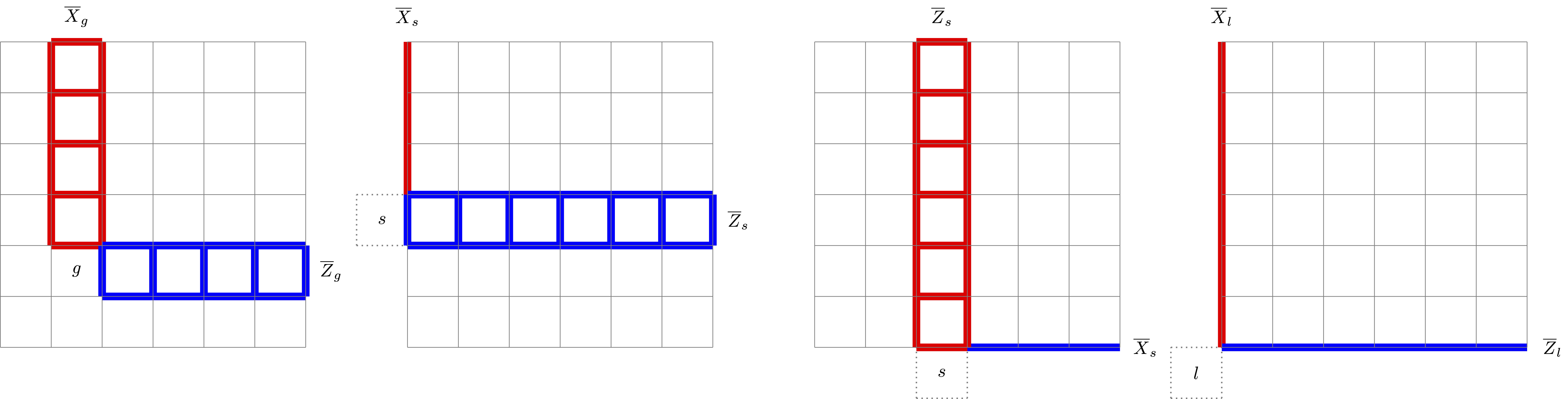}
\caption{Virtual qubit operators for the Bacon-Shor subsystem code. From left-right, we have respectively an example of a (i) gauge qubit, (ii) horizontal stabilizer qubit, (iii) vertical stabilizer qubit, and (iv) logical qubit. Note that while the gauge qubits are in one-to-one correspondence with the square plaquettes of the lattice, one can associate imaginary square plaquettes running along the left and bottom boundaries of the lattice for the latter three types of virtual qubits. Red and blue colors denote $X$ and $Z$ operators, both physical and logical, respectively. The depicted operators here generate the full Pauli group on $d^2$ qubits on a $d\times d$ lattice. The gauge group of the Bacon-Shor code is generated by the $\overline{X}_g$ and $\overline{Z}_g$ operators of the gauge qubits, the $\overline{Z}_s$ operators of the horizontal stabilizer qubits (second from the left above), and the $\overline{X}_s$ of the vertical stabilizer qubits (third from the left above), the latter two sets generating the stabilizer subgroup of the gauge group. The $\overline{X}_s$ operators of the horizontal stabilizer qubits and the $\overline{Z}_s$ operators of the vertical stabilizer qubits may be thought of as `destabilizers', as each detectable error will include a factor of such operators. Finally, $\overline{Z}_l$ and $\overline{X}_l$ are the logical operators (of the encoded logical qubit) of the Bacon-Shor subsystem code.}
\label{fig:bacon-shor-virtual-ops}
\end{figure*}

In stabilizer, or subspace, codes, we use the convention that the first $s$ virtual $\overline{Z}$ operators are the stabilizers, and identify the group they generate $\mathcal{S} = \langle \overline{Z}_1, \dots, \overline{Z}_{s} \rangle$ as the stabilizer group. Since these generators all mutually commute, they admit a simultaneous eigenbasis. Conventionally, an eigenbasis in which all the $\{ \overline{Z}_j\}_{j=1}^{s}$ take a value $+1$ is chosen to be the code space. Notably, $-\mathbb{I} \not\in \mathcal{S}$. Now, the centralizer of some arbitrary group $G$ in the Pauli group is defined as $\mathcal{Z}(G) = \lbrace p \; \vert \; p \in \mathcal{P}_{n}, pgp^{\dagger}=g \; \forall g \in G \rbrace$, the set of all Pauli operators that commute with every element of the group $G$. The centralizer of this stabilizer group is given by $\mathcal{Z}(\mathcal{S}) = \langle i\mathbb{I}, \overline{Z}_1, \dots, \overline{Z}_{s} , \overline{X}_{s+1}, \overline{Z}_{s+1}, \dots, \overline{X}_{n}, \overline{Z}_{n}\rangle$, where there is freedom as to how to choose the operators $\lbrace \overline{X}_{s+1}, \overline{Z}_{s+1}, \dots, \overline{X}_{n}, \overline{Z}_{n}\rbrace$ as long as they satisfy the Pauli commutation relations.

The centralizer contains the logical operators of this code, which are also the undetectable logical errors, in the subspace $\mathcal{L}(\mathcal{S}) = \mathcal{Z}(\mathcal{S}) \setminus \mathcal{S}$, where the notation $A \setminus B$ denotes the set of all the elements in $A$ that are not in $B$. The set of operators $\lbrace \overline{X}_{s+1}, \overline{Z}_{s+1}, \dots, \overline{X}_{n}, \overline{Z}_{n}\rbrace$ provides the logical operators for $k = n-s$ logical qubits. The smallest weight of any operator in $\mathcal{Z}(\mathcal{S}) \setminus \mathcal{S}$ gives the code distance $d$, where by weight we mean the number of physical qubits an operator non-trivially acts on. Detectable errors live in the space $\mathcal{P}_{n} \setminus \mathcal{Z}(\mathcal{S})$, and all such operators with weight $\leq\left\lfloor (d-1)/2 \right\rfloor$ are correctable. A code described in this manner is succinctly referred to as an $[[n,k,d]]$ stabilizer code.

Due to the abelian nature of stabilizer groups, and the fact that any pair of Pauli operators either commutes or anti-commutes, the normalizer of any stabilizer group $\mathcal{S}$, defined as $\mathcal{N}(\mathcal{S}) = \lbrace p \; \vert \; p \in \mathcal{P}_{n}, psp^{\dagger} \in \mathcal{S} \rbrace$, the set of all Pauli operators that leave the stabilizer group fixed upon conjugation, exactly coincides with the centralizer. However, for non-abelian groups, the normalizer and centralizer are generally different. While many expositions of stabilizer codes use the normalizer rather than the centralizer, when generalizing from stabilizer codes to subsystem codes, we need to be careful to choose the correct generalization which, as we will see, for subsystem codes is the centralizer. For this reason, we use the centralizer throughout the paper for consistency, even when discussing stabilizer codes.

In subsystem codes, we identify a non-abelian subgroup of the Pauli group $\mathcal{G} \subset \mathcal{P}_n$ as the gauge group. Due to its non-abelian nature, we now have $-\mathbb{I} \in \mathcal{G}$, unlike the case for stabilizer groups. (This implies that the normalizer of the gauge group is the entire Pauli group $\mathcal{N}(\mathcal{G}) = \mathcal{P}_{n}$; this is one of the reasons that for subsystems codes we must define relevant quantities in terms of the centralizer $\mathcal{Z}(\mathcal{G})$ rather than the normalizer.) The generators of this gauge group are typically chosen to be low-weight operators that can be measured with relative ease. But for a clearer theoretical picture of Bacon-Shor codes and, later, Floquet Bacon-Shor codes, we use a generating set of 
virtual qubit operators $\mathcal{G} = \langle i\mathbb{I}, \overline{Z}_1, \dots, \overline{Z}_s, \overline{X}_{s+1}, \overline{Z}_{s+1}, \dots, \overline{X}_{s+g}, \overline{Z}_{s+g}\rangle$.

The center of this gauge group, defined as the set of all elements of the gauge group that commute with all elements of the gauge group, is identified as the stabilizer subgroup $\mathcal{C}(\mathcal{G}) = \mathcal{Z}(\mathcal{G}) \cap \mathcal{G} = \mathcal{S}$ of the subsystem code, and is given in terms of the virtual qubit operators as $\mathcal{S} = \langle \overline{Z}_1, \dots, \overline{Z}_s\rangle$, where we define $1,\dots, s$ as the stabilizer virtual qubits. Meanwhile, the remaining non-trivial generators of the gauge group $\left\lbrace\overline{X}_{s+1}, \overline{Z}_{s+1}, \dots, \overline{X}_{s+g}, \overline{Z}_{s+g}\right\rbrace$ are the logical operators that define gauge degrees of freedom, which we refer to simply as (virtual) gauge qubits, numbered $(s+1), \dots, (s+g)$. The logical operators associated with these gauge qubits represent transformations that do not affect the codespace. Altogether, this gives us $s = \vert S \vert$ many stabilizer qubits, and $g = (\vert G \vert - \vert S \vert)/2$ many gauge qubits, where $\vert A \vert$ denotes the cardinality, or the number of independent generators, of some group $A$.

In the case of the nonabelian gauge groups for subsystem codes, the set of all operators $\mathcal{Z}(\mathcal{G}) \setminus S$ contains the bare logical operators for this subsystem code, which leave both the stabilizer as well as the gauge qubits intact, and only affect the logical qubits of the subsystem code.
However, owing to the gauge freedom in a subsystem code, we are also free to multiply these bare logical operators by any gauge operator, so that they can non-trivially operate on both the gauge and logical qubits, without impacting the state of the stabilizer qubits. The set of all such dressed logical operators operators is given by the subset of the centralizer of the stabilizer subgroup that excludes all elements in the gauge group, $\mathcal{L}(\mathcal{G})=\mathcal{Z}(\mathcal{S}) \setminus \mathcal{G}$. The weight of the smallest operator in the space $\mathcal{L}(\mathcal{G})$ gives the code distance $d$. As with stabilizer codes, detectable errors live in the space $\mathcal{P}_{n} \setminus \mathcal{Z}(\mathcal{S})$, and all operators in this space with weight $\leq\left\lfloor (d-1)/2 \right\rfloor$ are correctable. A code with gauge degrees of freedom described in this manner is referred to as an $[[n,k,g,d]]$ subsystem code, where $k = n - s - g$ is the number of logical qubits.

\subsection{Virtual qubit operators for the Bacon-Shor code}

The Bacon-Shor code is a prototypical subsystem code defined on an $L \times M$ square lattice.
The gauge group associated with this subsystem code is generated by all nearest-neighbor horizontal $XX$ checks and vertical $ZZ$ checks, i.e.
$\mathcal{G} = \left\langle X_{i,j} X_{i,j+1}, Z_{i,j}Z_{i+1,j} \mid {i \in [L], j \in [M]} \right\rangle$, where we use the notation $[K] = \lbrace 0, \dots, K-1 \rbrace$. The stabilizer subgroup of this gauge group is generated by operators that are either the product of all horizontal $XX$ checks along two neighboring columns, or the product of all vertical $ZZ$ checks along two neighboring rows.

The rank of this gauge group is simply the number of edges on the square lattice, $\vert G \vert = L(M-1) + M(L-1)$. Meanwhile, the rank of the stabilizer subgroup described above is given by $\vert S \vert = (L-1) + (M-1)$. This leaves us with $g = (\vert G \vert - \vert S \vert)/2 = (M-1)(L-1)$ many gauge qubits, which is exactly the number of square plaquettes in the $L \times M$ lattice. Thus, we can associate each gauge qubit with a square plaquette of the lattice. Since the number of physical qubits is given by $n = ML$, we are left with $k = n - g - s = 1$ logical qubit in this subsystem code.

The pair of bare logical operators for this subsystem code can be taken to be the product of $X$ operators along the left-most column, and similarly the product of $Z$ operators along the bottom-most row, which we choose to identify as the logical $\overline{X}$ and $\overline{Z}$ operators respectively. Up to stabilizer transformations, the logical $\overline{X}$ ($\overline{Z}$) operator is equivalent to the product of $X$ ($Z$) operators along any column (row).
Of course, these bare logical operators can also be multiplied by any gauge operator to yield (gauge) equivalent dressed logical operators.

We also identify the operators for the gauge qubits in this subsystem code, which we noted above can be associated with the square plaquettes of the lattice. We can take the $\overline{X}$ operator for a given gauge qubit to be the product of $XX$ checks on all the edges above the square plaquette, including the top edge of the square plaquette, associated with this gauge qubit. Similarly, the $\overline{Z}$ operator for this gauge qubit may be taken to be the product of $ZZ$ checks on all the edges to the right of its square plaquette, including the right edge of the square plaquette.

The identification of gauge qubits with square plaquettes also motivates the identification of the stabilizer qubits and the logical qubit with some part of the physical lattice. With the definitions of the stabilizer and logical qubit operators given above, it is intuitive to identify the horizontal $Z$-type stabilizer qubits with imaginary square plaquettes sharing right-side edges with the left-most column, and the vertical $X$-type stabilizer qubits with similar imaginary square plaquettes sharing top-side edges with the bottom-most row.
Although we noted earlier that the stabilizer generators of the code can be written as the logical $\overline{Z}$ operators of stabilizer qubits, one could equally well identify a subset of these generators as the logical $\overline{X}$ operators instead for those particular stabilizer qubits. Here, we adopt a convention that identifies the vertical stabilizers as the logical $\overline{X}$ operators of vertical stabilizer qubits, and the horizontal stabilizers as the logical $\overline{Z}$ operators of horizontal stabilizer qubits, as depicted in Fig. \ref{fig:bacon-shor-virtual-ops}.

Similarly, the logical qubit can be associated with an imaginary square plaquette with its top-right corner at the origin of the lattice residing in its bottom-left most corner. With such an identification of all virtual qubits with square plaquettes on the lattice, we denote e.g. $\overline{X}_{i,j}$ to be the logical $X$ operator of the virtual qubit associated with the square plaquette with its top-right corner at the $(i,j)$ coordinate of the $L \times M$ square lattice, where $i \in [M]$ and $j \in [L]$. This correspondence between virtual qubits and plaquettes is depicted in Fig.~\ref{fig:bacon-shor-virtual-ops}, where examples of the three types of virtual qubits (gauge, stabilizer, and logical) are shown. The explicit expressions for the logical operators of the various virtual qubits, in terms of the operators of the physical qubits, which reside on the nodes of the square lattice, is given in Eq.~\eqref{eqn:virtual-qubit-ops}.

\begin{widetext}
\begin{equation}
    \begin{alignedat}{5}
    &\text{Horizontal $Z$-type stabilizers:} &\qquad& \overline{Z}_{0,i} = \prod_{j=0}^{M-1} Z_{j,i} Z_{j,i-1} &\qquad & \overline{X}_{0,i} = \prod_{j=i}^{L-1} X_{0,j} &\qquad & (1 \leq i \leq L-1) \\
    &\text{Vertical $X$-type stabilizers:} &\qquad & \overline{Z}_{i,0} = \prod_{j=i}^{M-1} Z_{j,0}  &\qquad& \overline{X}_{i,0} = \prod_{j=0}^{L-1} X_{i,j} X_{i-1,j} &\qquad & (1 \leq i \leq M-1)\\
    &\text{Gauge qubits:} &\qquad& \overline{Z}_{i,j} = \prod_{k=i}^{M-1} Z_{k,j} Z_{k,j-1} &\qquad & \overline{X}_{i,j} = \prod_{k=j}^{L-1} X_{i,k} X_{i-1,k} &\qquad & (1 \leq i,j \leq M-1,L-1) \\
    &\text{Logical qubit:} &\qquad& \overline{Z}_{0,0} = \prod_{i=0}^{M-1} Z_{i,0} &\qquad & \overline{X}_{0,0} = \prod_{i=0}^{L-1} X_{0,i} &\qquad &
    \end{alignedat}
\label{eqn:virtual-qubit-ops}
\end{equation}
\end{widetext}

One way to fix a gauge of the Bacon-Shor code is to fix the values of some of the gauge qubit logical operators. Threshold behavior in such gauge fixings of the Bacon-Shor code have been previously explored in the context of 2D compass codes\cite{PhysRevX.9.021041}.

\section{Floquet-Bacon-Shor Code}
\begin{figure*}[!tp]
    \centering
\includegraphics[width=\columnwidth]{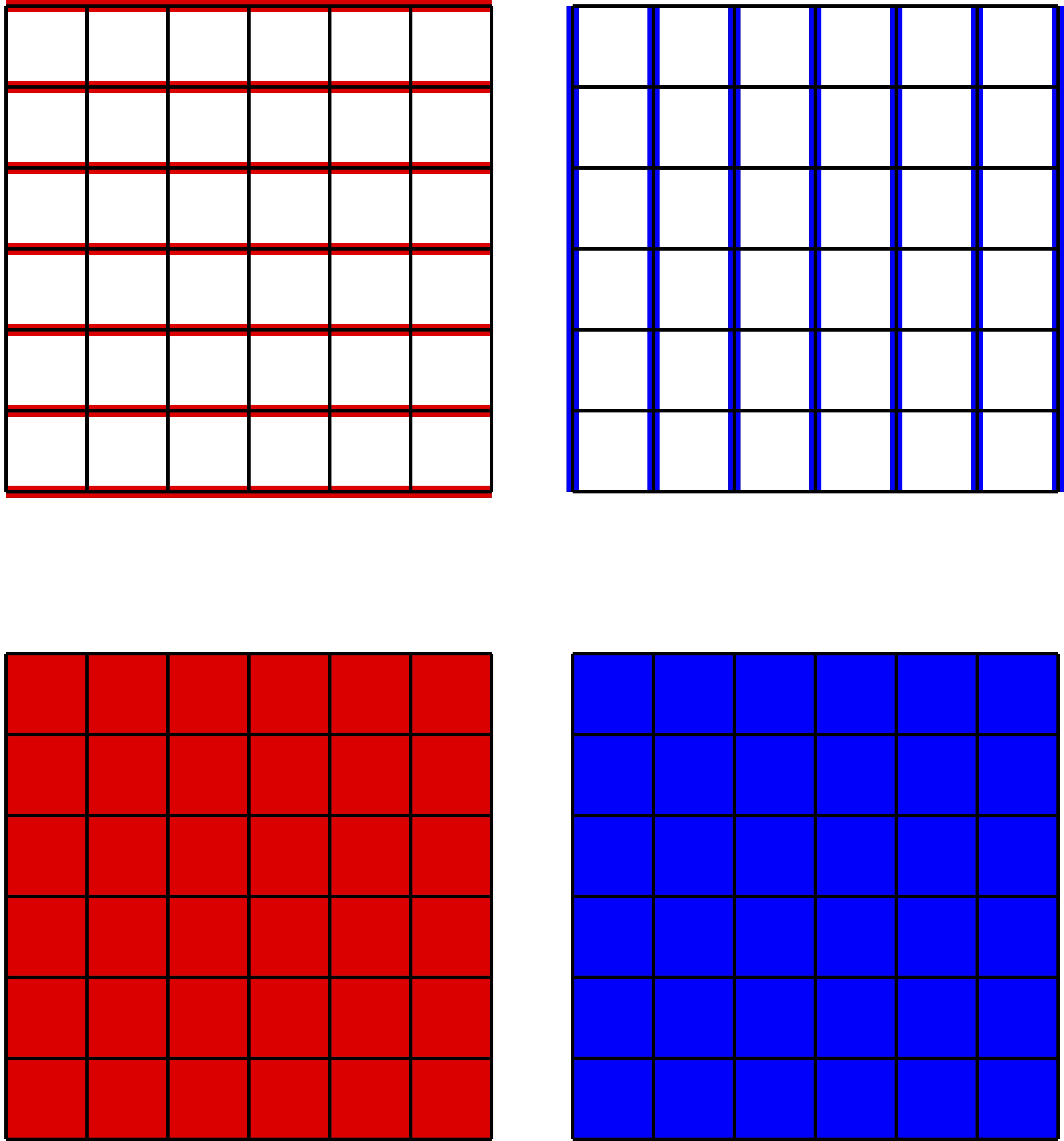}
\caption{(Top) A simple period 2 measurement schedule for the Bacon-Shor subsystem code, given by measuring all the horizontal $XX$ checks (colored red) and therefore all the $X$-type (vertical) stabilizers, followed by measuring all the vertical $ZZ$ checks (colored blue) and similarly all the $Z$-type (horizontal) stabilizers. \\
(Bottom) From a Floquet perspective, this corresponds to moving between two instantaneous stabilizer groups (ISGs) that are total gauge fixings of the parent subsystem code, where we fix all the $\overline{X}$ operators of the gauge qubits in the first round, followed by fixing all their $\overline{Z}$ operators in the second round.
Each cell, or equivalently a gauge qubit, colored either red or blue corresponds to gauge fixing its $\overline{X}$ or $\overline{Z}$ operator respectively.}
\label{fig:bacon-shor-measurement-schedule}
\end{figure*}

In order to measure the stabilizers of the Bacon-Shor subsystem code, it suffices to use a period 2 measurement schedule, such that we measure all the horizontal $XX$ checks in the first round, followed by all the vertical $ZZ$ checks. From a Floquet perspective, this corresponds to periodically moving between two different gauge fixings of the Bacon-Shor code, one where the $\overline{X}$ operators of all the gauge qubits has been fixed, followed by fixing all the $\overline{Z}$ gauge qubit operators, as depicted in Fig. \ref{fig:bacon-shor-measurement-schedule}. We can view either gauge fixing as a stabilizer code that contains the Bacon-Shor stabilizers as a subset, but also includes either the $\overline{X}$ or $\overline{Z}$ gauge qubit operator. It does not matter whether these gauge qubit operators are fixed to $\pm 1$, and in general they will take random values upon each measurement, with the only constraint being the values of the stabilizers.

In order to free up space for an additional logical qubit, we must unfix one of the gauge degrees of freedom. In general, such an unfixing, or gauge defect, is difficult to do while ensuring all the Bacon-Shor stabilizers also get measured. However, as we show here, it is possible to maintain such a gauge defect in a dynamical sense, where the encoded space for the corresponding logical qubit changes across measurement rounds, as seen in previously constructed Floquet codes.

\subsection{Measurement schedule and ISGs}
\begin{figure*}[!tp]
    \centering
\includegraphics[width=\linewidth]{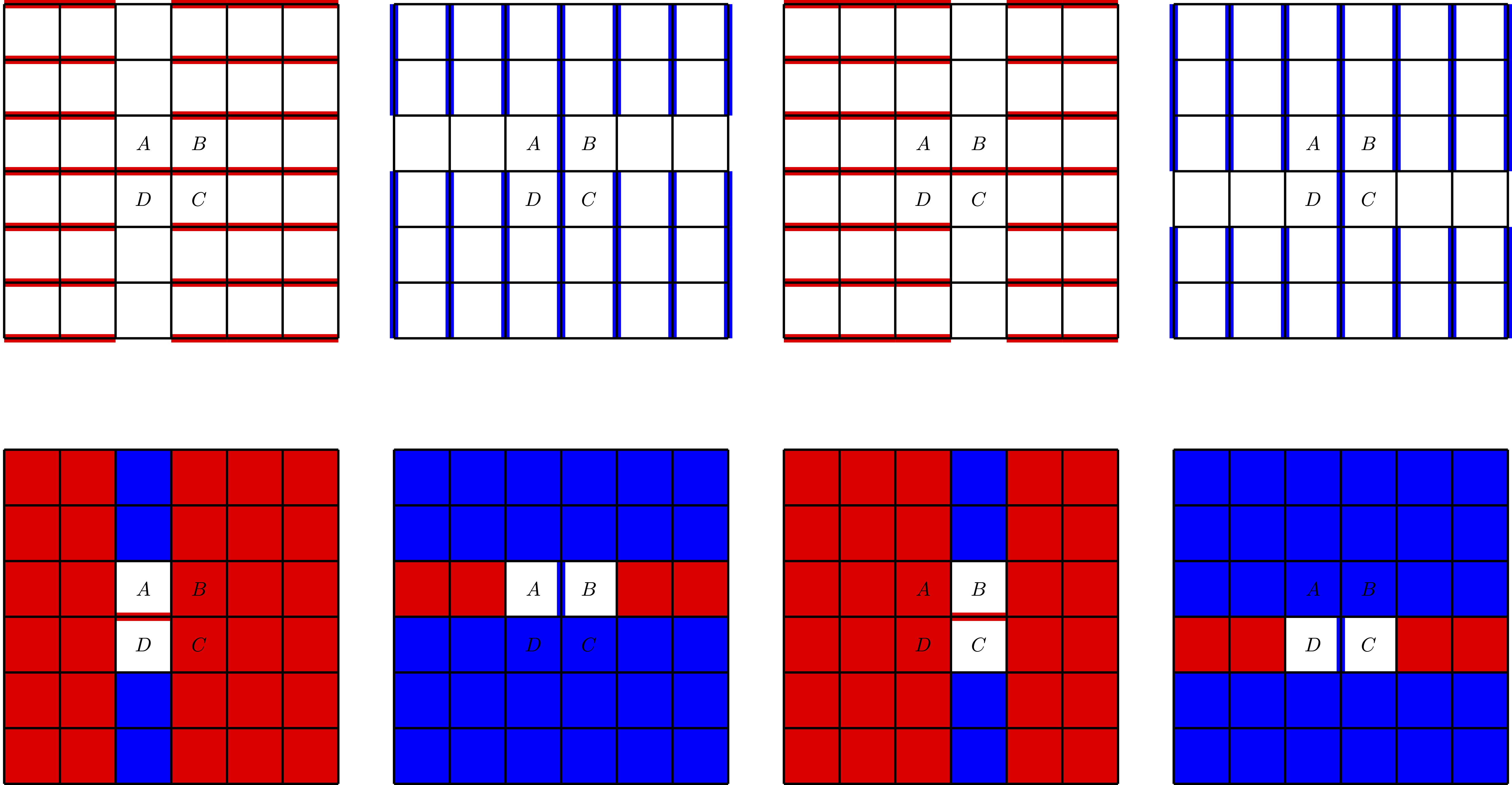}
\caption{Measurement schedule (top) with period 4 that maintains a gauge defect across the entire measurement cycle, thereby realizing a dynamical logical qubit in addition to the usual Bacon-Shor logical qubit. Horizontal $XX$ checks are colored red, while vertical $ZZ$ checks are colored blue. After one complete period of this measurement schedule, the induced instantaneous stabilizer groups (ISGs) (bottom) achieve steady state. In addition to the usual Bacon-Shor stabilizers, each of the ISGs contains the measured check operators, as well as any elements from the previous ISG that commute with the currently measured checks. Any cell, or equivalently a gauge qubit, colored either red or blue corresponds to gauge fixing its $\overline{X}$ or $\overline{Z}$ operator respectively.}
\label{fig:measurement-schedule}
\end{figure*}

We can adopt the period 4 measurement schedule shown in the top part of Fig.~\ref{fig:measurement-schedule}. The core idea here is that we distribute the measurement of the Bacon-Shor stabilizers into four, instead of two, rounds, allowing us to prevent one of the gauge degrees of freedom from being fixed, thereby introducing a gauge defect that evolves across the measurement rounds and serves as the dynamical logical degree of freedom.

In order to describe the measurement process, it is helpful to label four of the square plaquettes that serve a special role in identifying the dynamical logical qubit. For simplicity, we will consider a $d \times d$ square lattice. As in Section \ref{secn:bacon-shor}, we label the location of square plaquettes with the location of their top-right corner. Since square plaquettes are in one-to-one correspondence with gauge qubits, we will use the two terms interchangeably. For odd $d$, the gauge qubit A is identified as the plaquette at location $\left( \frac{d-1}{2}, \frac{d+1}{2} \right)$, B at $\left( \frac{d+1}{2}, \frac{d+1}{2} \right)$, C at $\left( \frac{d+1}{2}, \frac{d-1}{2}\right)$, and D at $\left( \frac{d-1}{2}, \frac{d-1}{2} \right)$. Similarly, for even $d$, A is identified as the plaquette at $\left( \frac{d}{2} - 1, \frac{d}{2} + 1\right)$, B at $\left( \frac{d}{2}, \frac{d}{2} + 1\right)$, C at $\left( \frac{d}{2}, \frac{d}{2}\right)$, and D at $\left( \frac{d}{2} - 1, \frac{d}{2}\right)$. In either case, these four plaquettes all share a single corner that lies roughly at the center of the square lattice. We also denote columns AD and BC, and rows AB and CD as the unique columns and rows containing the obvious choice of plaquettes, as well as edges AB, BC, CD and AD as the edges between the respective plaquettes.

A complete period of the measurement cycle consists of 4 measurement rounds, which are subsequently repeated. In each measurement round, we measure either all $XX$ or $ZZ$ checks except along a particular column or row of plaquettes, that we refer to as the defect column or row. The defect column is column AD at round 0, and column BC at round 2, while the defect row is row AB at round 1, and row CD at round 3. In the defect column or row, we measure a single $XX$ or $ZZ$ check on the edge that shares the same label as the column. For instance, at round 0, we measure all the $XX$ checks except those in column AD, in which which we only measure the $XX$ check on edge AD. With this measurement schedule, the Bacon-Shor stabilizers associated with the defect columns and rows get measured only once in every period, while all other stabilizers get measured twice. The gauge defect is identified with the pair of gauge qubits that straddle the unique edge in the defect column or row that we measure a check operator on. It evolves from the pair AD at round 0, to AB at round 1, to BC at round 2, to CD at round 3, and then again to AD, repeating the cycle. In essence, it is this gauge defect that serves as the dynamical logical qubit. Note that an $XX$ ($ZZ$) check on any edge equals the product of $\overline{X}$ ($\overline{Z}$) operators for the two square plaquettes sharing that edge, fixing which removes precisely one gauge degree of freedom.

After one complete period, the instantaneous stabilizer groups (ISGs) induced by the measurement schedule settles into a steady state, and is shown in the bottom part of Fig.~\ref{fig:measurement-schedule}. Each of the four ISGs contains all the check operators measured in the current round, or equivalently all the virtual gauge qubit $\overline{X}$ ($\overline{Z}$) operators in the plaquette columns (rows) along which we measure all the horizontal (vertical) $XX$ ($ZZ$) checks. Along the defect plaquette column (row), the ISG consists of $\overline{Z}$ ($\overline{X}$) operators for all the gauge qubits except those that straddle the edge whose $XX$ ($ZZ$) check we measure in the current round. For this pair of gauge qubits, the gauge defect, we fix only the product of their $\overline{X}$ ($\overline{Z}$) operators. In addition, all the ISGs also contain the Bacon-Shor stabilizers. The fixed gauge qubit opertors randomly take either value $\pm 1$ each time they are measured afresh. Only a subset of these operators survive across rounds, and can therefore serve as error syndromes.

For example, the logical $\overline{X}$ operators of the gauge qubits along row CD (apart from the gauge qubits at cells C and D) are initialized at round 2, commute with all measurements at round 3, and are re-measured at the next round, which is round 0 of the next measurement cycle, at which point they can serve as syndromes. 
Note that when such an operator is initialized at round 2, its measurement outcome at this round is inherently random. However, whatever value it measures to be remains fixed through the next round.  In the round after that, its re-measurement should, in the absence of errors, deterministically produce the same result as 2 rounds ago when it was initialized, at which point it can serve as a detector or a syndrome.
In a similar way, the logical $\overline{X}$ operators of the gauge qubits along rows AB serve as syndromes at round 2, while the $\overline{Z}$ operators of the gauge qubits along columns AD and BC serve as syndromes at rounds 1 and 3 respectively. In each of these cases, the operators are initialized 2 rounds before they are re-measured at the round at which they serve as syndromes. In addition to these gauge fixed operators, the Bacon-Shor stabilizers also serve as error syndromes just as they do in the parent subsystem code.

Let $\mathcal{S}^{(r)}$ denote the ISG at round r. As an example of the description above, $\mathcal{S}^{(4k)}$ ($k \geq 1$) contains the $\overline{X}$ operators of all the gauge qubits except those that belong in the missed plaquette column containing gauge qubits A and D. This ISG contains the $\overline{Z}$ operators of all the gauge qubits in this column except the gauge qubits A and D. These $\overline{Z}$ operators were measured and have persisted from the previous round, and commute with all other elements of the current ISG. In addition, the product $\overline{X}_{A} \overline{X}_{D}$, which is simply the $XX$ check measured on the edge shared between qubits A and D, also exists as an element of the ISG $\mathcal{S}^{(4k)}$.

Using the definitions of the gauge qubits A, B, C and D described above, and depicted in Fig.~\ref{fig:measurement-schedule}, the pair $(\overline{X}, \overline{Z})$ of logical operators for the dynamical qubit evolves as $(\overline{X}_A, \overline{Z}_A \overline{Z}_D)$ at round 0, $(\overline{X}_A \overline{X}_B, \overline{Z}_B)$ at round 1, $(\overline{X}_B, \overline{Z}_B \overline{Z}_C)$ at round 2, and  $(\overline{X}_C \overline{X}_D, \overline{Z}_C)$ at round 3. We discuss how this choice preserves logical information from one round to the next further below.

In any ISG, only gauge fixed $\overline{Z}$ operators appear anywhere in the column(s), and similarly only gauge fixed $\overline{X}$ operators appear anywhere in the row(s), in which the gauge defect belongs. 
Suppose, as an example, that there was a gauge fixed $\overline{X}_g$ operator in the column AD for some gauge qubit $g$ located above gauge qubit $A$ at round 0. The operator $\overline{X}_A \overline{X}_g$ is now also a logical operator, but one that has lower weight than $\overline{X}_A$, since it is now composed of the product of all $XX$ checks between cells $A$ and $g$. In particular, if cell $g$ was located a constant distance away from cell $A$, then the ISG at round 0 would have constant code distance. This argument can be straightforwardly generalized to all the other rounds, and for the $\overline{Z}$ operators by considering gauge fixed $\overline{Z}$ operators in the same row as the gauge defect(s).
Thus, the absence of gauge fixed $\overline{X}$ operators in the same column(s) or $\overline{Z}$ operators in the same row(s) as the gauge defects ensures that the code distance of any of the ISGs is maintained at $\sim d$ on a $d \times d$ square lattice. Of course, this also relies on the vertex shared by the plaquettes A, B, C and D lying roughly $d/2$ away from any boundary of the lattice, as the code distance would shrink if the gauge defect was instead brought closer to any lattice boundary.

The logical operators for the Bacon-Shor qubit do not change across these ISGs, and are still given by the operators $\overline{X}_{0,0}$ and $\overline{Z}_{0,0}$ from Eq.~\eqref{eqn:virtual-qubit-ops}. In the honeycomb Floquet code \cite{Hastings2021dynamically}, there are similar static logical operators arising from non-trivial homological loops. These were referred to as inner logical operators since they belong to the center of the parent gauge group, though the measurement schedule prevents them from ever being measured. Here, the logical operators for the Bacon-Shor qubit are not inner logical operators in the sense that they do not belong to the center of the gauge group. So instead, we simply refer to them as static logical operators, even though of course, the equivalence class of such operators up to instantaneous stabilizer transformations does evolve from round to round.

The logical operators for the dynamical qubit introduced as a result of the measurement schedule introduced above do change across rounds. Roughly speaking, they correspond to the virtual qubit operators associated with the gauge defect that arises from fixing the product of logical operators, but not the individual operators themselves, of a pair of neighboring gauge qubits among $A, B, C$ and $D$, and that evolves from round to round. Whenever we fix the product of virtual qubit operators of a particular type, $\overline{X}$ or $\overline{Z}$, of such neighboring gauge qubits, the logical operator of the same type for the dynamical qubit is given by the virtual qubit operator for either gauge qubit, while that of the opposite type is given by the product of the virtual qubit operators of the two gauge qubits. 

In order to preserve logical information across these four ISGs, it is necessary that some representative, up to instantaneous stabilizer operations, of a logical operator at round $r$ persists as a logical operator in the next round $r+1$. This is formalized in Eq.~\eqref{eqn:preserve-info}. Roughly speaking, this condition ensures that if $\ket{\psi^{(r)}} = \alpha \ket{0^{(r)}} + \beta \ket{1^{(r)}}$ is the state of the dynamical logical qubit at round $r$, then the measurement schedule projects it to $\ket{\psi^{(r)}} \rightarrow \alpha \ket{0^{(r+1)}} + \beta \ket{1^{(r+1)}} = \ket{\psi^{(r+1)}}$ at round $r+1$, where the codewords 
$\ket{0^{(r)}}$ etc change across the ISGs, but the pair of coefficients $\alpha$ and $\beta$, and therefore the logical information, does not. Specifically, this condition says that if $S^A$ and $S^B$ are two stabilizer groups associated with stabilizer codes $A$ and $B$ respectively, with associated logical operators $O^A$ and $O^B$, where $O^j \subset \mathcal{N}(S^j)\setminus S^j$, then if there exist $s^{A} \in S^{A}$ and $s^{B} \in S^{B}$ such that
\begin{equation}
s^A O^A = s^B O^B
\label{eqn:preserve-info}
\end{equation}
for each pair of $O^A$ and $O^B$, then logical information is preserved. This is also discussed more in the appendix.

Following Eq.~\eqref{eqn:preserve-info}, we must identify pairs of stabilizer elements $s^{(r)}, s^{(r+1)} \in \mathcal{S}^{(r)}, \mathcal{S}^{(r+1)}$ that connect a logical operator for the ISG $\mathcal{S}^{(r)}$ to one for the ISG in the next round $\mathcal{S}^{(r+1)}$ in order to preserve the logical information contained in the dynamical qubit across these rounds. These pairs are given in Table~\ref{table:logical-ops}. The dynamical logical operators for each ISG are depicted in Fig. \ref{fig:gauges-logicals}.


\begin{figure*}[!t]
    \centering
\includegraphics[width=\linewidth]{gauges_logicals.jpg}
\caption{Dynamically evolving logical operators for a single dynamical qubit in the Floquet-Bacon-Shor code. The physical support of the logical $\overline{X}$ ($\overline{Z}$) operator consists of the vertices of the square lattice lying along the red (blue) lines. For instance, at round 0, the logical $\overline{X}$ operator is given by the product of $XX$ checks above cell/gauge qubit $A$, while the logical $\overline{Z}$ operator is given by the product of physical $Z$s along the two blue lines drawn, which also equals the product of all $ZZ$ checks to the right of cells/gauge qubits $A$ and $D$. A topologically equivalent (via instantaneous stabilizer elements) $\overline{X}$ operator at round 0 is shown in dotted red lines.}
\label{fig:gauges-logicals}
\end{figure*}

\bgroup
\def\arraystretch{1.5}
\begin{table}[!h]
\begin{center}
\begin{tabular}{ |c | c | c | c | c |}
\hline
 Round & 0 & 1 & 2 & 3 \\
 \hline\hline
 $\overline{X}^{(r)}$ & $\overline{X}_A$ & $\overline{X}_A \overline{X}_B$ & $\overline{X}_B$ & $\overline{X}_C \overline{X}_D$ \\
 \hline
 $\overline{Z}^{(r)}$ & $\overline{Z}_A \overline{Z}_D$ & $\overline{Z}_B$ & $\overline{Z}_B \overline{Z}_C$ & $\overline{Z}_C$ \\
 \hline\hline
 $s_{x}^{(r)}$ & $\overline{X}_A \overline{X}_D \cdot \overline{X}_C$ & $\mathbb{I}$ & $\overline{X}_A$ & $\mathbb{I}$ \\
 \hline
 $s_{x}^{(r-1)}$ & $\mathbb{I}$ & $\overline{X}_B$ & $\mathbb{I}$ & $\overline{X}_B \overline{X}_C \cdot \overline{X}_D$ \\
 \hline
 $s_{z}^{(r)}$ & $\mathbb{I}$ & $\overline{Z}_D \cdot \overline{Z}_A \overline{Z}_B$ & $\mathbb{I}$ & $\overline{Z}_B$ \\
 \hline
 $s_{z}^{(r-1)}$ & $\overline{Z}_C \overline{Z}_D \cdot \overline{Z}_A$ & $\mathbb{I}$ & $\overline{Z}_C$ & $\mathbb{I}$ \\
 \hline
\end{tabular}
\caption{\label{table:logical-ops}Evolution of the dynamical logical operators $\overline{X}^{(r)}$ and $\overline{Z}^{(r)}$ across the four measurement rounds, where the locations of the gauge qubits/plaquettes A, B, C and D are described in the main text. The stabilizer element pairs $s_{j}^{(r)}$ and $s_{j}^{(r-1)}$ for type $j$ connect the logical operators across subsequent rounds, $s_{x}^{(r)} \overline{X}^{(r)} = s_{x}^{(r-1)} \overline{X}^{(r-1)}$, and similarly for the $\overline{Z}^{(r)}$ operators. Note that in this equation, each of the operators $\overline{X}^{(r)}$, $s_{x}^{(r)}$ and $s_{x}^{(r-1)}$ are specified by their values given in the column for round $r$, while $\overline{X}^{(r-1)}$ is given by its value under column $r-1$. Note in particular that the value of $s_{x}^{(r-1)}$ under column $r$ is not in general equal to that of $s_{x}^{(r)}$ under column $r-1$. The dynamical logical operators are depicted in Fig. \ref{fig:gauges-logicals}.}
\end{center}
\end{table}
\egroup

\subsection{Multiple Dynamical Qubits}
\label{subsecn:multiple-dynamical-qubits}
Introducing additional dynamical logical degrees of freedom amounts to adding more gauge defects to the measurement schedule. It is straightforward to generalize from the construction above for a single dynamical logical qubit. Consider relabeling the set of plaquettes $(A, B, C, D)$ described above as $(A_0, B_0, C_0, D_0)$, and identifying another such similar set of plaquettes $(A_1, B_1, C_1, D_1)$ where we repeat the measurement schedule identified above. Thus, for instance, we would now have two defect columns, $A_0 D_0$ and $A_1 D_1$, at round 0. The code distance would now be proportional to the shorter of the two distances from any edge of the square lattice to either of the two resultant gauge defects. To make one of these distances the same for both gauge defects, we can put them in the same plaquette column or row. Suppose, without loss of generality, that we put them in the same row. The optimal choice would be to place both gauge defects roughly $d/3$ distance away from each other, as well as either vertical edge of a $d \times d$ square lattice, but roughly $d/2$ away from the horizontal edges. Since this already reduces the code distance to $\sim d/3$, one might as well add two more gauge defects such that all of these lie roughly $d/3$ units away from any edge of a $d \times d$ square lattice.

\begin{figure*}[!tp]
    \centering
\includegraphics[width=\linewidth]{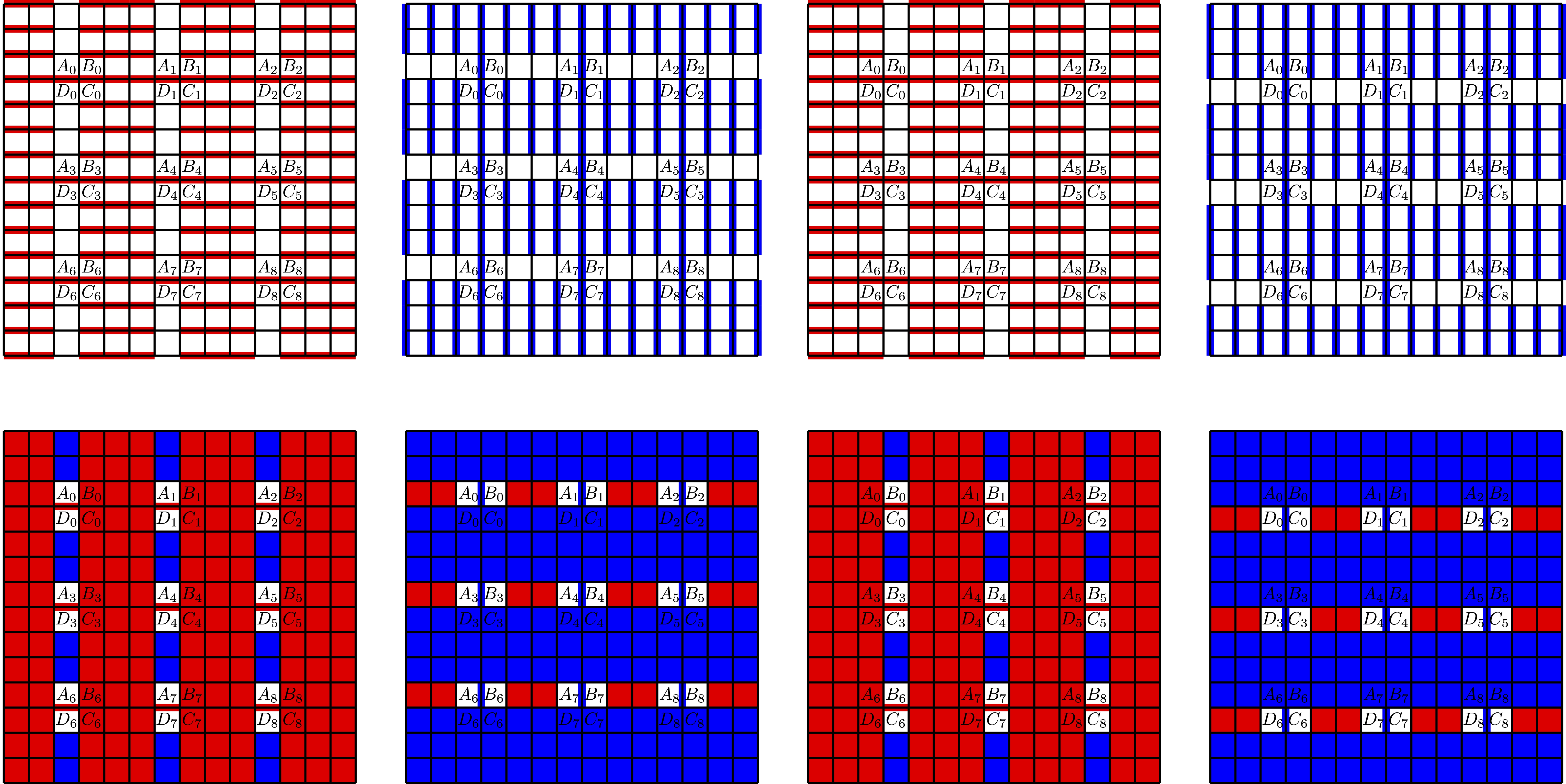}
\caption{Measurement schedule (top) for hosting 9 dynamical logical qubits in addition to the usual Bacon-Shor subsystem logical qubit. The sequence of instantaneous stabilizer groups (ISGs) this measurement schedule induces (below) maintains the introduced gauge defects throughout the measurement cycle.}
\label{fig:multiple-logical-qubits}
\end{figure*}

Generalizing this construction, we see that while one could in principle keep adding more dynamical logical qubits by simply adding more gauge defects, the overall code distance decreases by the same amount as one goes from $m^2$ to any number between $m^2 + 1$ and $(m+1)^2$ dynamical logical qubits, where $m$ is an integer. The optimal placement for $m^2$ many gauge defects is such that we place $m$ of these gauge defects in the same row and column, such that the boundaries of the resultant defect lattice are roughly $d/(m+1)$ away from the edges of the square lattice. Therefore, adding $k$ dynamical logical qubits to the Bacon-Shor code reduces the code distance to roughly $d/(\sqrt{k} + 1)$, or $\Theta(d/\sqrt{k})$. This situation is depicted in Fig. \ref{fig:multiple-logical-qubits} for the illustrative case of 9 dynamical logical qubits.

Note that the usual Bacon-Shor code defined on an $\sqrt{n} \times \sqrt{n}$ lattice with $n$ physical qubits has code distance $\sqrt{n}$ and admits a single logical qubit. Meanwhile, the bound for local 2D subsystem codes relating these parameters is $kd = O(n)$ \cite{PhysRevA.83.012320}. In the case of the Floquet-Bacon-Shor code family introduced here, we can choose the number of dynamical gauge defects to be a constant fraction of the total number of physical qubits, which would also simultaneously set the code distance to be a constant. In other words, we would have $k = \Theta(n)$ and $d=\Theta(1)$, which does saturate the bound $kd = O(n)$, unlike the parent Bacon-Shor subsystem code. Alternatively, we can choose $k = \Theta(\sqrt{n})$ and $d=\Theta(\sqrt{n})$, which would also saturate the $kd = O(n)$ bound.

In the presence of biased noise, one may instead want to adjust the $X$ and $Z$ distances, $d_X$ and $d_Z$, separately. For instance, in the hypothetical case of infinite $Z$ biased noise, one could keep adding more dynamical gauge defects along a particular column, which progressively shortens $d_X$ but maintains the same value of $d_Z$, as all these dynamical qubits lie the same distance away from the vertical boundaries of the square lattice. Alternatively, as with the Bacon-Shor parent subsystem code, we may elongate the horizontal dimension of the square lattice itself while keeping the vertical dimension fixed.

\section{Error Detection and Correction}

Here, we describe how errors can be detected and corrected in the Floquet-Bacon-Shor codes introduced above. We will first outline some general principles in identifying the operators to form detectors out of, the structure of the resultant decoding graph, as well as a calculation of an upper bound to the effective code distance using a method from \cite{fu2024errorcorrectiondynamicalcodes}. Finally, we will provide numerical results using Stim \cite{gidney2021stim} and PyMatching \cite{Higgott2025sparseblossom}.

In the usual Bacon-Shor subsystem code, an odd number of $X$ errors in any row are gauge equivalent to a single $X$ error anywhere in that row. Similarly, an odd number of $Z$ errors in any column are, up to multiplication by gauge group elements, equivalent to a single $Z$ error anywhere in that column. A chain of such errors will anti-commute with, and therefore get detected by, the stabilizers that reside at the boundaries of such chains. For instance, a single $X$ error on any qubit on the lattice will anti-commute with the horizontal $Z$-type stabilizers that straddle the row in which this $X$ error occurred. Error correction then proceeds by applying a single correcting $X$ or $Z$ operator anywhere in the relevant row or column. Conventionally we can choose the left-most column to apply correcting $X$ operators along, and the bottom-most row to apply correcting $Z$ operators on. All such errors of weights $< d/2$ can be corrected in this manner on a $d \times d$ square lattice.

In the Floquet-Bacon-Shor code described above, we have additional stabilizers that protect the encoded dynamical logical qubits. These are given by the gauge fixed operators that lie along the defect columns and rows of the various ISGs. We will refer to these as transient stabilizers, to distinguish them from the usual Bacon-Shor stabilizers that run along the length of the entire lattice, which we shall refer to as permanent stabilizers. As we discuss below, a decoding graph can be built up entirely from transient stabilizers, so that standard matching decoding algorithms can be applied. The collection of such gauge operators, or transient stabilizers, in any defect column or row form a domain wall across which single qubit errors cannot propagate, since we skip the measurements along that column or row. For instance, assuming temporarily a single dynamical logical qubit for simplicity, at round $4k$ ($k \geq 1$), we do not measure any of the $XX$ checks in the defect column AD except on the edge AD, so that single qubit $X$ errors are equivalent, up to instantaneous stabilizer operations, to any single $X$ error on either side of the domain wall in the column AD, depending on which side of this wall the error occurred, except along the row that contains the edge AD. This means that unlike the Bacon-Shor code, an even number of $X$ errors, for instance, can bring us out of the dynamical codespace if we have an odd number of $X$ errors on both sides of the domain wall.

\subsection{Decoding and Code Distance}
For simplicity, we will consider the case of a single dynamical logical qubit; the the extension to multiple logical qubits is straightforward, as briefly discussed towards the end of this section. In addition, since each of the ISGs is equivalent to a clockwise $\pi/2$ rotation of the previous ISG, together with interchanging $X \leftrightarrow Z$, it suffices to discuss errors occurring between any two consecutive rounds.

We have a choice with respect to a decoding approach. We could form detectors out of the permanent Bacon-Shor stabilizers along with the gauge fixed operators along the defect rows/columns. This approach, however, in general leads to a decoding hyper-graph, which is typically more complicated to decode then decoding graphs. To see that this approach leads to a hypergraph, consider the ISG at round $4k$ and a single qubit $X$ error anywhere to the right of the defect column AD away from the gauge defect or the lattice boundary. Any such error would flip both the pair of permanent stabilizers and the gauge fixed operators that straddle the row along which this error occurred. Thus, such an error would lead to a hyper-edge in the decoding hyper-graph that connects all $4$ of these stabilizers.

\begin{figure*}[!t]
    \centering
\includegraphics[width=\linewidth]{detector_example_1q.jpg}
\caption{An example pair of detectors formed through initialization at round 0 and re-measurement at round 2, in the plaquette column immediately to the left of column AD. The top detector can be expressed as the product of measured $\overline{X}$ operators of the gauge qubit $g$ at rounds 0 and 2, i.e. $\overline{X}_{g}^{(0)} \cdot \overline{X}_{g}^{(2)}$, or alternatively as the product of highlighted measured checks shown at those two rounds. The bottom detector can similarly be expressed as the product of measured $\overline{X}$ operators of the gauge qubit $g$ and the vertical stabilizer qubit $s$ shown above at rounds 0 and 2, i.e. $\overline{X}_{g}^{(0)} \overline{X}_{s}^{(0)} \cdot \overline{X}_{g}^{(2)} \overline{X}_{s}^{(2)}$, or alternatively as the product of highlighted measured checks shown.}
\label{fig:detector-example-1q}
\end{figure*}

To avoid creating a hypergraph, a useful strategy is to consider the transient stabilizers together with the product of permanent and transient stabilizers in the same plaquette row/column as the error syndromes.
For instance, at round $4k$ $(k \geq 1)$, for any square cell/plaquette along the plaquette row AB, except cells A and B, we would take the product of all $X$ checks above this as one stabilizer syndrome and the product of all $X$ checks below this cell as another stabilizer syndrome. The former is the virtual $\overline{X}$ operator of the gauge qubit associated with that plaquette, whereas the latter is the product of the $\overline{X}$ operators of this gauge qubit and the vertical stabilizer qubit along the respective column. An example of a pair of such transient stabilizers forming detectors is given in Fig. \ref{fig:detector-example-1q}. In this example, they serve as syndromes between round $4k$, when they are initialized, and $4k+2$, when they are re-measured, persisting through the intermediary round $4k+1$. Single-qubit $Z$ errors on any column, except the left or right most columns or those bordering either cell A or B, will flip a nearest-neighbor pair of such transient stabilizers. A single-qubit $Z$ error on either the left or right most column will flip a single such transient stabilizer, depending on which side of the domain wall it occurred. We also have permanent $X$-type stabilizers along plaquette columns BC and AD, which are measured at rounds $4k$ and $4k+2$ respectively, and similarly permanent $Z$-type stabilizers along plaquette rows AB and CD, which are measured at rounds $4k+1$ and $4k+3$ respectively. Detectors are only formed out of these permanent stabilizers by comparison to the measurement of such stabilizers from 4 rounds ago, the same round in the previous measurement cycle.

\begin{figure*}[!t]
    \centering
\includegraphics[width=\linewidth]{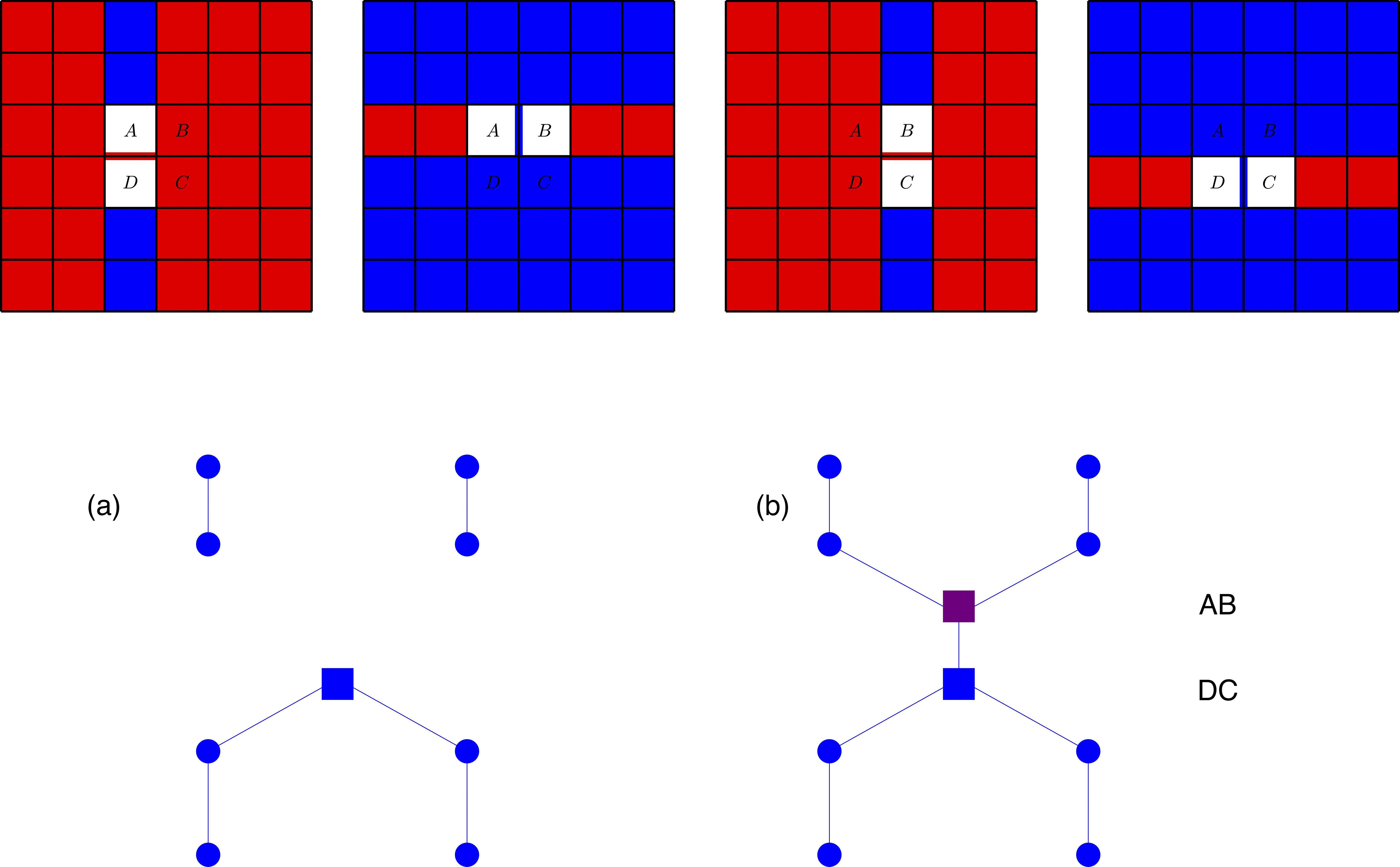}
\caption{(Top) ISG progression reproduced for convenience. Below, we depict the subgraph of the decoding graph that can detect errors occurring between rounds $4k+1$ and $4k-1$. All nodes represent $Z$ type stabilizers. Circular nodes depict temporary stabilizers that get initialized and measured at round $4k-1$, and then subsequently re-measured to form detectors and then destroyed at round $4k+1$. Square nodes represent the permanent Bacon-Shor stabilizers, which get measured once per measurement cycle. (a) Decoding using only the information available at the current time slice will necessarily exclude one of the permanent Bacon-Shor stabilizers, resulting in a disconnected decoding graph, and reducing the weight to $\sim d/2$, the length of either one of the straight line sub-graphs shown above. (b) By taking into account syndrome information received 2 rounds later, one can form edges in the decoding graph between the permanent Bacon-Shor stabilizer measured along row AB at round $4k+3$ (colored in violet), and the rest of the stabilizers. Edges depicted above between the violet and blue nodes cut across time slices separated by 2 measurement rounds. Note that $X$ errors occurring on the top and bottom most rows of the lattice, and similarly $Z$ errors occurring on the right and left most columns of the lattice, are not depicted in the figures above, and they would only flip a single detector residing on the boundary.}
\label{fig:decoding-graph}
\end{figure*}

One decoding strategy is to detect and correct errors immediately, in the same time slice as they are detected. In our case, this will lead to a smaller effective code distance, as we discuss shortly, though it may generally be an attractive strategy for Floquet codes in order to shorten the time between which syndrome bits are gathered and the decoder produces its output, as discussed later below.
To see the limitation of this approach in our case however, consider round $4k + 1$, when in addition to measuring the transient $Z$ stabilizers along every plaquette row except AB and CD, we also measure the permanent $Z$ stabilizer along row CD. This approach would result in a disconnected decoding graph with two components, one connecting the measured stablizers above the plaquette row AB and the other connecting the measured stabilizers below the plaquette row AB.
In this approach, however, an error chain of weight roughly half the lattice length would produce a trivial syndrome, resulting in a code distance of only $\sim d/2$. Such a chain could be distinguished from the single $X$ error if one also includes the $Z$-type stabilizer along plaquette row AB, but this does not get measured until measurement round $4k+3$. 

This observation suggests that a better decoding strategy would make use of information from multiple time slices. Observe that a single qubit $X$ error on any (vertex) row between the two plaquette rows AB and CD will flip the two permanent $Z$ stabilizers associated with these rows, separated by 2 measurement rounds. If we include such edges in the decoding graph, we can correct errors of weight $\sim d/2$, thereby restoring the code distance to $\sim d$. In any case, all single qubit errors lead to defects in either one or two stabilizers, and therefore standard matching techniques can be applied for the decoding task. The structure of the decoding graph is illustrated in Fig. \ref{fig:decoding-graph}.

Numerical simulations carried out in a simplified code capacity error model show that the code distance does indeed scale as $\sim d$, as discussed in greater detail in Section \ref{subsecn:numerical-results}. The code distance we report here is the smallest weight of any non-trivial operator that produces a trivial syndrome, given the decoding graph, which in turn depends on the error model described above. Computing the distance of a code is generally NP hard, and we expect it remains so for Floquet codes. Moreoever, the notion of distance for a Floquet code is more nuanced than for a stabilizer or even subsystem codes. To begin with, the code distance is not simply the minimum code distance of any of the ISGs considered as stabilizer codes in isolation, as the time dynamics may allow for small weight errors to evolve into logical errors, even if the ISGs have large code distances by themselves.

Additionally, we can multiply instantaneous logical operators not only by instantaneous stabilizer elements, but also elements of the ISGs before and after the current round. This is so because an error of the form $\mathcal{L}^{(r)} s^{(r-1)}$ between measurement rounds $r-1$ and $r$, where $\mathcal{L}^{(r)}$ is a logical operator at round $r$, and $s^{(r-1)} \in \mathcal{S}^{(r-1)}$, has the same effect as an error of the form $\mathcal{L}^{(r)}$ at round $r$. Similarly, an error of the form $s^{(r+1)}\mathcal{L}^{(r)}$ between measurement rounds $r$ and $r+1$ will also have the same effect. In the present case, such transformations do not affect the code distance scaling reported here. One way to see this is to note that the union of the measurements in two consecutive rounds of measurement forms a subsystem code. The minimum distance of all such subsystem codes across the entire measurement cycle provides a better upper bound to the distance of a Floquet code, than the minimum ISG distance, but in the general case it remains only an upper bound.

The above two upper bounds to the distance of a Floquet code are described more formally in \cite{fu2024errorcorrectiondynamicalcodes} (which appeared a day after the initial posting of this paper on the arXiv preprint server), where they are referred to as the ISG and subsystem distances respectively. 
While the ISG distance informs us of the maximum distance if we only took into account the current ISG, the subsystem distance offers a better bound on the distance of the dynamical code since it also incorporates the effect of the next ISG in the sequence. An even better bound for the dynamical code distance than the ISG or subsystem distances can be obtained if we extended this window even further out into the measurement schedule, such that it extends out to all ISGs that come later in the measurement schedule in which a subset of elements of the current ISG can still persist. This notion was formalized in \cite{fu2024errorcorrectiondynamicalcodes} as the unmasked distance.

The unmasked distance depends on the notion of unmasked stabilizers, which are roughly speaking the stabilizers from any given ISG that survive a given measurement sequence, and are able to form detectors. In the case of the Floquet-Bacon-Shor codes we have described, the distance algorithm of \cite{fu2024errorcorrectiondynamicalcodes} identifies precisely such stabilizers, and predicts the same code distance $\sim d$ that we argued for above using the decoding graph perspective.

In the case of the ISG at round(s) $4k$, these detectors are formed by the $\overline{X}$ ($\overline{Z}$) operators of all the gauge qubits along the plaquette row AB (column AD), except for gauge qubits A and B (A and D), as well as their products with the permanent Bacon-Shor stabilizers in the same row (column), along with the permanent Bacon-Shor stabilizers along column AD and row AB. Meanwhile, the gauge group consists of all other gauge fixed $\overline{X}$ operators at round $4k$ as well as the gauge $\overline{Z}$ operators that destabilize the former in the next round $4k+1$. We provide a detailed calculation using the distance algorithm of \cite{fu2024errorcorrectiondynamicalcodes} applied to an example Floquet-Bacon-Shor code in Appendix \ref{appdx:code-distance}.

As the authors state, the distance algorithm of \cite{fu2024errorcorrectiondynamicalcodes} only works in the case of perfect measurements.
We present some initial numerical results and discuss complications that may arise when considering measurement errors in Section \ref{subsecn:numerical-results}.
Moreover, even without measurement errors, the algorithm may not be able to capture the propagation of low-weight errors into logical errors such as in the planar Floquet codes constructed in \cite{vuillot2021planar}, leading to constant distance.
Although it is beyond the scope of this current work to develop an algorithm that fully encompasses notions of code distance in Floquet or dynamical codes, it may be fruitful to consider the space-time connectedness of low weight errors for such a task. Generally speaking, we may say that two Pauli operators $\sigma_{\alpha}$ and $\sigma_{\beta}$ are space-time connected in a dynamical code if their difference is a chain of instantaneous stabilizer elements, i.e. $\sigma_{\alpha} = s_{r+t} \dots s_{r} \sigma_{\beta}$. It may be the case that a low weight Pauli error $\sigma_{\beta}$ traverses, in its spacetime path, an undetectable logical error, before evolving into a partially detectable error, whose correction does not correct the previous logical error, or even gets absorbed into the ISG. Such a logical error occurs in \cite{vuillot2021planar}, for instance. It may therefore be desirable to keep the number of time slices between which stabilizers are measured and used as syndromes as minimal as possible when designing Floquet codes. This intuition partially motivated the distance $\sim d/2$ decoding strategy we described above.

We also note that like the parent Bacon-Shor subsystem code, the Floquet-Bacon-Shor family of codes introduced here do not possess a threshold. The issue is that the weight of the stabilizers used in both examples of codes grows as the distance increases, so that erroneous stabilizer measurements get likelier with increasing distance for any fixed value of independent error probability.

\subsection{Self-correcting errors}
Suppose that a 2-qubit $XX$ error occurs on an edge somewhere above cell A between rounds 0 and 1 mod 4. This will flip the two transient, but not the two permanent, $Z$-type stabilizers that straddle this edge, lie along column AD, and are measured in round 1 mod 4. In the very next round 2 mod 4, we carry out a measurement of an $XX$ check operator on precisely the same edge, and the error becomes part of the ISG. In other words, the appropriate correction operator automatically applies itself because of the measurement schedule itself. This occurs even if this 2-qubit error anti-commutes with a logical operator at round 0 mod 4, such as if it occurred on the top edge of cell A. We show this more explicitly in the appendix. Such self-correcting errors may be understood as a residual gauge degree of freedom in the Floquet version of the parent subsystem code, as even though such errors do not belong in the ISG and will produce syndromes, they are nevertheless gauge checks of the parent subsystem code.

Furthermore, note that a single qubit $X$ error occurring between rounds 0 and 1 mod 4 on any one row on say the right side of the blue domain wall AD is ISG equivalent to a single $X$ error on the right boundary of the domain wall. It cannot be transported further left due to the existence of the domain wall, which in turn exists because we do not carry out any check measurements along the plaquette column of this wall. However, two measurement rounds later, we do carry out the measurement of this check and the error can propagate further left. In turn, this implies that whether a single qubit $X$ error occurs on the left or the right of the blue domain wall, the correction operator two measurement rounds later is to apply a single $X$ operator somewhere on the left of the instantaneous blue domain wall, which has now shifted one column to the right and exists along plaquette column BC.

\subsection{Multiple Dynamical Qubits}
The discussion above generalizes fairly straightforwardly to the case of adding multiple dynamical logical qubits. In this case, the main difference is that there are now more temporary stabilizers used to form detectors in each plaquette row and column. In particular, there is one such detector between any neighboring pair of gauge defects, as well as a detector between any boundary gauge defect and the lattice boundary as before. An even number of $X$ errors between gauge defects in the same plaquette row are now part of the ISG at round 0, and no longer count as errors. Only single qubit $X$ errors between any consecutive pair of gauge defects, or a gauge defect and a lattice boundary count as errors, and similarly for $Z$ errors. It was already noted in Section \ref{subsecn:multiple-dynamical-qubits} that the minimum distance of all ISGs shrinks to roughly $d/\sqrt{k}$ with the addition of $k$ dynamical qubits. The arguments for the single logical qubit discussed above for the code distance generalize to the case of multiple logical qubits as well, and therefore it continues to be the case that the smallest weight of any error operator leading to a trivial syndrome is roughly $d/\sqrt{k}$.

\subsection{Numerical Results}
\label{subsecn:numerical-results}

\begin{figure*}[!t]
    \centering
\includegraphics[width=0.8\linewidth]{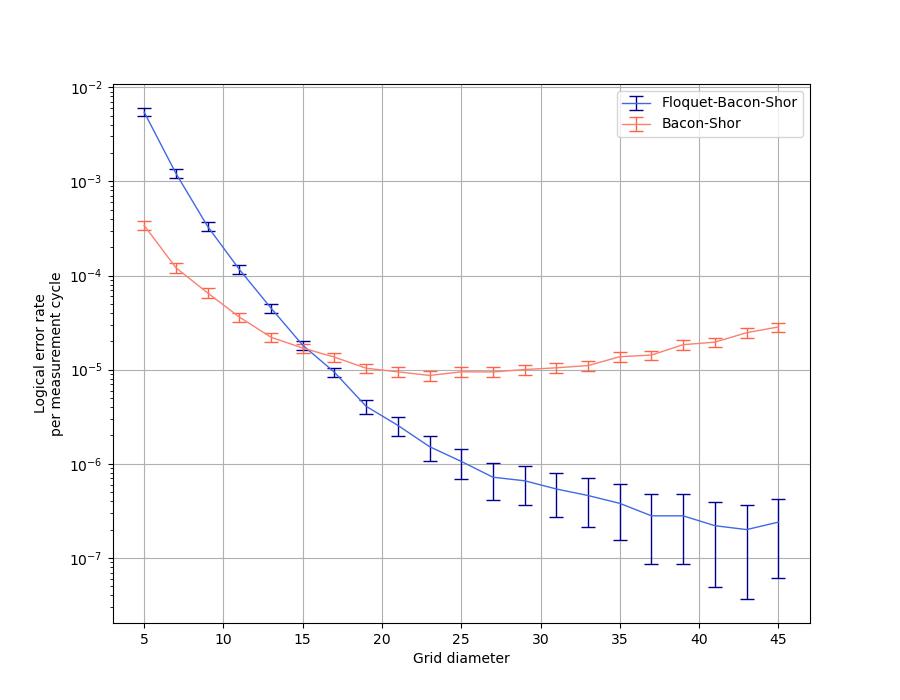}
\caption{Numerical simulations results for the usual Bacon-Shor code and Floquet-Bacon-Shor code with a single additional dynamical qubit, under a simple code capacity error model, with independent depolarizing noise channels on each qubit with error rate $5 \times 10^{-3}$ before each measurement round. We run each circuit for 50 measurement cycles, and sample $10^6$ shots or 500 logical errors, whichever comes first. The reported logical error rate is the total number of logical error per measurement cycle per sampled shot. The error bars are estimated as described in the main text.}
\label{fig:code-capacity}
\end{figure*}

\begin{figure*}[!t]
    \centering
\includegraphics[width=0.8\linewidth]{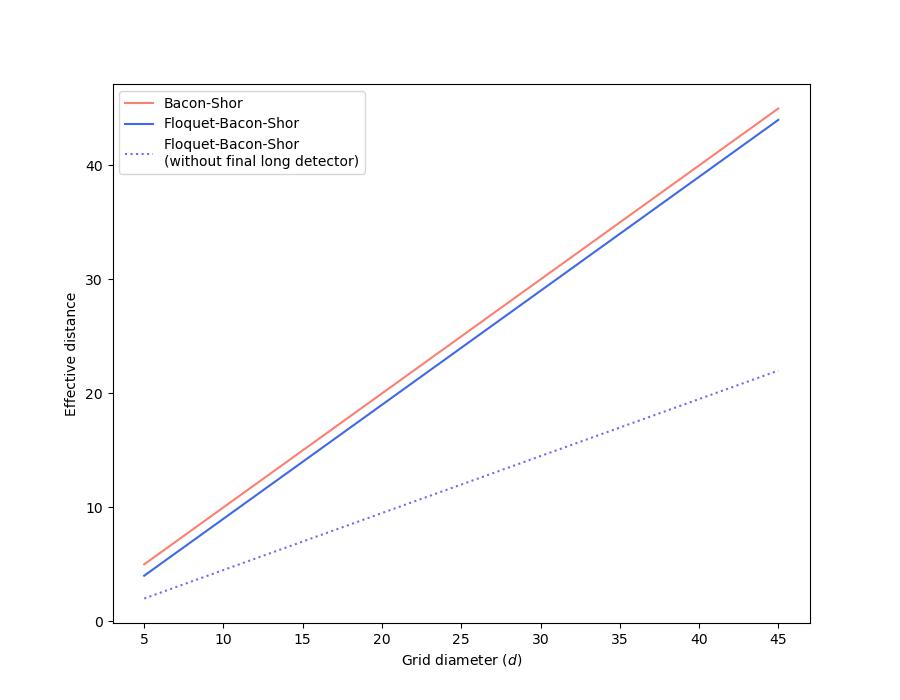}
\caption{Effective distance computed through Stim's $\texttt{shortest\_graphlike\_error}$ under the simplified code capacity model described in the main text. While the Bacon-Shor (BS) has effective distance equal to the grid diameter ($d$) in this noise model, the Floquet-Bacon-Shor (FBS) code with a single dynamical qubit has effective code distance scaling as $(d-1)$. In both cases, we form detectors out of the last measured checks (the $Z$-type checks of the fourth round in Fig. \ref{fig:measurement-schedule} in the case of the FBS code, and similarly $Z$-type checks of the second round in Fig. \ref{fig:bacon-shor-measurement-schedule} in the case of the BS code) using individual qubit measurements in the computational basis. Additionally, for the FBS code, we form a detector out of the product of the last measured value of the permanent stabilizer along row CD (which is unmeasured in the last round, but measured two rounds before that). If we skip forming this detector, then the distance scales as $\left\lfloor \frac{d-1}{2} \right\rfloor$ instead.}
\label{fig:code-capacity-distance}
\end{figure*}

Here, we numerically benchmark our code construction using Stim \cite{gidney2021stim} to simulate the error correction circuits and the implementation of minimum weight perfect matching provided by PyMatching \cite{Higgott2025sparseblossom} as our decoder. We assume a simple code capacity error model, in which single-qubit depolarizing noise is applied to all qubits before each round of measurements, and assume perfect measurement outcomes for each of the measured checks. The results of a numerical simulation under this noise model are shown in Fig. \ref{fig:code-capacity}, and details provided in Appendix \ref{appdx:numerics}. We observe that under this simple noise model with a physical error rate of $5 \times 10^{-3}$, the Floquet-Bacon-Shor code with a single additional logical qubit starts yielding better logical error rates than the usual Bacon-Shor around $d=15$, the size of the square grid diameter. We conjecture that this is so because the Floquet-Bacon-Shor code creates more detectors than the usual Bacon-Shor code, owing to the existence of temporary, or transient, stabilizers. For both types of codes, we observe that the logical error rate stops improving beyond a certain grid diameter, and even starts to worsen, more obviously so for the Bacon-Shor code up to the values of the grid diameter we tested. In turn, this is so because neither code possesses a threshold, even in this simple code capacity noise model, due to the growing weight of the stabilizers (both transient and permanent) with the size of the lattice, or the grid diameter.

The error bars shown in Fig. \ref{fig:code-capacity} are estimated as follows. We consider the binomial variable that specifies whether a logical error occurred or not once a circuit is sampled. We define a logical error for the Floquet-Bacon-Shor code as an error in either the dynamical $Z$ operator of the single additional dynamical logical qubit, or the $Z$ operator of the static logical qubit inherited from the parent subsystem code. We also define a measurement cycle as one complete period of measurement rounds as shown in Fig. \ref{fig:measurement-schedule}. We estimate the probability of obtaining a logical error per sampled circuit as simply the total number of sampled logical errors divided by the total number of shots as $\hat{p}$. The estimated standard deviation of this binomial variable is then given by $\sqrt{\hat{p} (1 - \hat{p}) / N_{shots}}$. In our simulations, we report the logical error rate per cycle, defined as the total number of logical errors sampled divided by the product of the number of shots $N_{shots}$ and the number of measurement cycles $N_{cycles}$, i.e. $p = \hat{p}/N_{cycles}$. In turn, this provides the estimated standard deviation of the logical error rate per cycle as $(1/N_{cycles})\sqrt{\hat{p} (1 - \hat{p}) / N_{shots}}$. The error bars in Fig. \ref{fig:code-capacity} are $99\%$ percent confidence intervals, of length $2.576$ times this standard deviation, around the data points. In our simulations, we sampled $10^6$ shots, or $5 \times 10^{2}$ logical errors, whichever came first.


The scaling of the effective code distance under this simplified code capacity noise model, shown in Fig. \ref{fig:code-capacity-distance}, is computed using the $\texttt{shortest\_graphlike\_error}$ function in Stim \cite{gidney2021stim}. This function, as the name suggests, reports the smallest weight operator, composed of errors that flip at most two detectors, that does not itself produce a detection event. Since our choice of detectors produces a decoding graph (and not a hyper-graph), such an error coincides with the smallest weight logical error, and its weight therefore provides the effective code distance. The effective distance of the usual Bacon-Shor code scales as $d$ on a $d \times d$ square lattice, and the Floquet-Bacon-Shor with a single logical qubit has code distance scaling roughly the same as $(d-1)$. This is so because the defect is always placed $\left\lfloor \frac{d-1}{2} \right\rfloor$ edges away from the nearest lattice boundary, and each edge contributes a weight of 2 to the logical operator, which therefore has weight $2\left\lfloor \frac{d-1}{2} \right\rfloor$, which equals $d-1$ for odd $d$ and $d-2$ for even $d$. Note that placing the defect anywhere else in the lattice would lead to a lower value for the effective distance.

We form detectors as described in the main text for the two cases. In both cases, we additionally form detectors out of the last set of measured checks, which in both cases are $Z$-type checks, by measuring all qubits in the computational ($Z$) basis. In the case of the Floquet-Bacon-Shor code, the permanent $Z$-type stabilizer along row CD is not measured in the last round, but we implicitly re-measure this stabilizer when we measure all qubits in the $Z$ basis after this last round. Therefore, we can also form a final detector for the stabilizer along row $CD$ if we compare to its last previous measurement in the second round of the last measurement cycle. If we instead skip forming this long detector, then the effective distance becomes $\left\lfloor \frac{d-1}{2}\right\rfloor$. This is so because errors of weight $\sim d/2$ running along the length of some column at the very final round will flip this detector but will leave all other detectors unflipped, a situation that is schematically similar to the one depicted in Fig. \ref{fig:decoding-graph}. We provide more specific details in Appendix \ref{appdx:numerics}.

\begin{figure*}[!t]
    \centering
\includegraphics[width=0.8\linewidth]{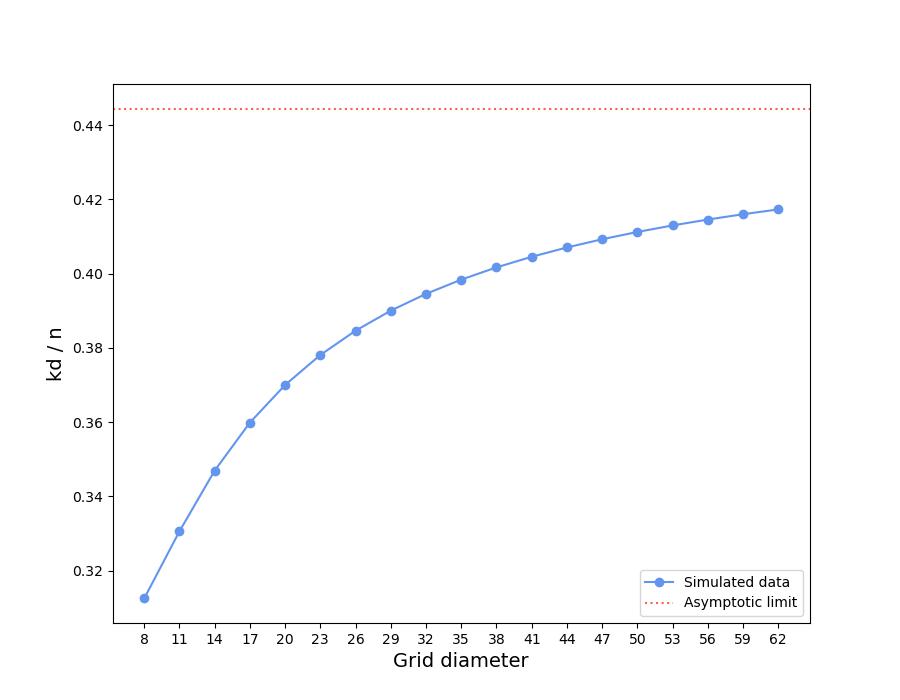}
\caption{Numerical evidence of the Floquet-Bacon-Shor code family saturating the subsystem BPT bound $kd = O(n)$. We parametrize the grid diameter as $L = 3q+2$, so that the total number of physical qubits is $n = (3q+2)^{2}$, where $q$ is allowed to increase from $q=2$ to $q=20$. The number of introduced dynamical qubits is $q^2$ so that the total number of logical qubits is $k = q^{2} + 1$, including the logical qubit of the parent subsystem code. We numerically find $d=4$ for all instances of the code family, and plot the ratio $kd/n$. In the large $L$ (alternatively, $q$) limit, this ratio approaches the asymptotic value $4/9$.}
\label{fig:kdn-scaling}
\end{figure*}

Finally, we provide numerical evidence that the Floquet-Bacon-Shor code family can saturate the subsystem BPT bound $kd = O(n)$ \cite{PhysRevA.83.012320}. Using the same code capacity error model as described above, we increase the number of dynamical logical qubits as $k = \Theta(n)$ while maintaining a constant distance $d=O(1)$. We do so by introducing as many gauge defects such that neighboring defects in the same column at round 0, for example, are only one square plaquette away from each other, or the lattice boundary (in the case of the top or bottom most defects in that column), while each of the neighboring defect columns are at most two square plaquettes away from each other, or the lattice boundary in the case of the left or right most columns. A single square plaquette translates to two edges, so that we expect an effective code distance of 4 in all such cases. We verify this through numerical simulations using Stim \cite{gidney2021stim}. In particular, we parametrize the lattice length, or the grid diameter, as $L = 3q+2$, and let $q$ steadily increase in integer values. The number of introduced logical dynamical qubits, or the number of gauge defects, equals $q^2$, and therefore the total number of logical qubits is $k = q^2 + 1$, since we also include the logical qubit of the parent subsystem code. A logical error occurs when there is an undetected flip in any of the $k$ logical $Z$ operators. Using Stim's $\texttt{shortest\_graphlike\_error}$ function, we find an effective code distance of $4$ in all cases. For any given value of the parameter $q$, we have $kd/n = 4(q^2 + 1)/(3q+2)^2$, which asymptotes to the constant $4/9$ as $q\rightarrow \infty$. We numerically observe in Fig. \ref{fig:kdn-scaling} that our code family approaches this asymptotic constant with increasing value of $q$, or alternatively the grid diameter, thereby providing numerical evidence in addition to the previous arguments that the Floquet-Bacon-Shor code family can saturate the subsystem bound $kd = O(n)$, unlike the parent Bacon-Shor subsystem code. 
Note that if the code family did not saturate the subsystem BPT bound, then the ratio $kd/n$ would asymptote to zero instead in the limit of large $n$, or equivalently in the limit of large grid diameter. For instance, for the Bacon-Shor code family, $kd/n \sim 1/\sqrt{n}$, which goes to $0$ for large $n$.

The noise model we described above is admittedly quite simplistic. A more realistic noise model would, in addition to single-qubit depolarizing noise on all qubits before each round of measurements, also feature measurement errors. In order to benchmark our code family under such a noise model, we introduce both reset and measurement errors. After the qubits are intialized in the $\ket{0}$ state, they are all subject to single-qubit bit flip channels. Immediately afterwards, they are subject to single-qubit depolarizing noise channels as in the simplified code capacity noise model described earlier, before the first set of measurements. In addition to these pre-measurement round depolarizing channels, the measurements of the check operators are themselves subject to errors such that they read out to the wrong value with some probability. In our simulation described below, we took the probability of noise in the depolarizing noise channels, the post-reset bit-flip channels, as well as the measurement error probability to all be $10^{-3}$, and sampled either $10^7$ shots, or $10^3$ logical errors, whichever came first. We note that while this uniform noise model is more realistic than the simplified code capacity model described earlier, it is not reproducing circuit level details, since e.g. 2q depolarizing noise channels after every (erroneous) check measurement may be more valid, since these measurements involve a 2-qubit gate, and moreover not all checks at a particular round can be carried out in the same time step, since most of them involve qubits participating in more than one check. Here, we only benchmark our code family in somewhat simplified noise models, and leave more accurate modeling of noise for future studies.

\begin{figure*}[!t]
    \centering
\includegraphics[width=0.8\linewidth]{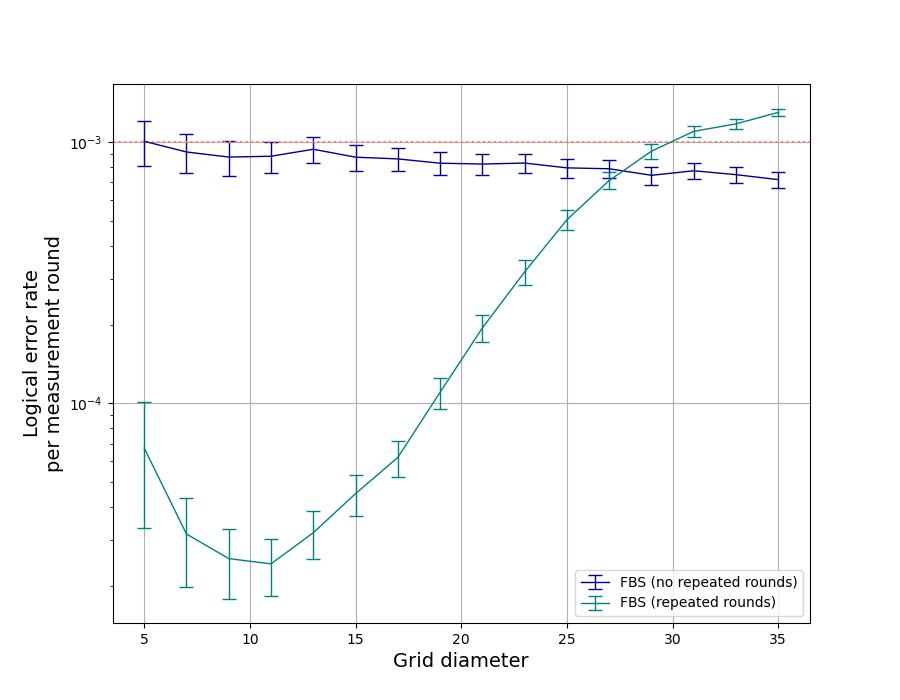}
\caption{Numerical results in the presence of measurement, reset and pre-measurement single-qubit depolarizing errors. With no repeated rounds, we carry out $O(d)$ many measurement cycles, each cycle consisting of the 4 measurement rounds shown in Fig. \ref{fig:measurement-schedule}. With repeated rounds, we essentially carry out a single measurement cycle, but where each of the 4 rounds is repeated $O(d)$ many times before the next measurement round is carried out. More details are provided in Appendix \ref{appdx:numerics}.}
\label{fig:fbs-repeated-rounds}
\end{figure*}

To ensure fault tolerance in the presence of measurement errors, we would need to repeat each measurement round $O(d)$ many times on a $d \times d$ lattice, not unlike the case of the usual Bacon-Shor code. However, there is now an important difference that we can either repeat the entire measurement cycle $O(d)$ times, or each round itself $O(d)$ many times. Schematically, a schedule where the measurement cycle is repeated $M$ times may be denoted $(S_0 S_1 S_2 S_3)^{M}$, whereas a schedule where each measurement round itself is repeated $M$ times is denoted $(S_0^M S_1^M S_2^M S_3^M)$. In the usual Bacon-Shor code, it should not make much difference whether we repeated the rounds or the cycles, as all the stabilizers that are used to form detectors persist across all rounds. The reason this makes a difference in the Floquet-Bacon-Shor code is that our detectors do not persist across all 4 of the ISGs. For instance, some $X$ type detectors are initialized at round 0 and then re-measured at round 2, but do not persist beyond that. In order to obtain reliable values of these detectors in the presence of measurement errors, it is reasonable to conjecture that we ought to take $O(d)$ repetitions in a period where they would have deterministic values.

This is borne out by our numerical simulations. The performance of the Floquet-Bacon-Shor code family under this noise model is depicted in Fig. \ref{fig:fbs-repeated-rounds}, with details provided of the numerical simulation in Appendix \ref{appdx:numerics}. For both cases, each measurement round is carried out $O(d)$ many times, where $d$ is the grid diameter. In the case of the repeated rounds version, where we carry out $O(d)$ rounds and only a single cycle, it is more reasonable to divide the logical error rate (per shot) by the number of measurement rounds (which are $O(d)$) and not the number of measurement cycles (which is just 1, in this case). To make a fair comparison, we also report the logical error rate per round (not cycle) for the usual measurement schedule. Note that we previously reported the logical error rate per measurement cycle, so that the values reported in Fig. \ref{fig:code-capacity} under the code capacity noise model would have to be further divided out by a factor of 4 in order to make a fair comparison with the results of Fig. \ref{fig:fbs-repeated-rounds}.

We observe from Fig. \ref{fig:fbs-repeated-rounds} that for low values of the grid diameter, the repeated rounds schedule outperforms the usual schedule in logical error rate per measurement round. As noted before, we expect this on the basis of being able to fault-tolerantly extract syndrome bits in the presence of measurement errors. Beyond a certain grid diameter of roughly $30$ however, the performance starts to worsen to the point that we do not get logical error rates per round below that of the physical error rate. In the repeated rounds schedule, we are also allowing for physical errors to build up to logical errors since a particular repeated measurement round, for example round 0, only checks for $Z$ errors, so that $X$ errors can build up during these rounds to form logical errors. The larger the grid distance, the larger the number of measurement rounds in this scheme, which may allow for a greater chance of logical errors occurring.

We note that the two versions we numerically benchmarked here are two extremes of a range of possibilities. For instance, instead of repeating each measurement round only once in a cycle and repeating many cycles, or repeating each measurement round many times in a cycle and carrying out only one cycle, one may want to divide up the total number of rounds $N_r$ such that each measurement round is repeated $j > 1$ many times in each cycle for a total of $N_r / j$ many cycles. This may allow a better balancing between allowing a fault-tolerant extraction of syndrome bits in the presence of measurement errors while taking care not to allow physical errors that remain undetected for a certain number of rounds to build up to a logical error. There may therefore exist better choices of dividing up the total number of measurement rounds into the number of repeated rounds and the number of measurement cycles, and we leave this as an open question for future studies.

\section{Conclusions}
In this paper, we have described the introduction of dynamical logical qubits to the Bacon-Shor subsystem code by modifying the measurement schedule of the weight 2 check operators that generate the gauge group of the parent sub-system code. Such a measurement schedule introduces and maintains gauge defects throughout the measurement cycle. This broad prescription of identifying the virtual qubit degrees of freedom in a subsystem code, and picking a measurement schedule that can maintain gauge defects across measurement cycles as a way to introduce dynamical logical qubits may be more broadly applicable in constructing more examples of Floquet codes.

We note that an alternative measurement schedule such as the one discussed in this paper can significantly increase the number of logical qubits the code allows for. While the parent Bacon-Shor subsytem code hosts a single logical qubit, the Floquet-Bacon-Shor family of codes can allow $k = \Theta(n)$ many logical qubits, with the tradeoff of a having a constant distance $d=\Theta(1)$. This implies that the Floquet version can saturate the subsystem bound $kd=O(n)$ \cite{PhysRevA.83.012320}, even though the parent subsystem code does not.

We showed that such an alternative measurement schedule also introduces temporary stabilizers that have smaller weight than the permanent stabilizers of the parent Bacon-Shor subsystem code, but that can nevertheless be used to form detectors and perform error correction using standard matching techniques. We reported the code distance as the minimum weight of any error operator that would yield a trivial syndrome, and noted certain obstacles to forming an all-encompassing notion of Floquet code distance, especially in the fault-tolerant setting. We leave a detailed consideration of fault tolerance, together with fault tolerant implementation of logical gates, in this code to future work.

Lastly, we also note that skipping measurements of some check operators in some rounds was also used in \cite{gidney2023bacon} to introduce a threshold to the Bacon-Shor code. In our paper, we used such skipped measurements to introduce dynamical logical qubits to the Bacon-Shor code. We leave it as an open problem to investigate whether one could combine both sets of ideas to introduce a threshold to the Bacon-Shor code in the presence of dynamical logical qubits.

\section{Acknowledgments}
This material is based upon work supported by the U.S. Department of Energy, Office of Science,
National Quantum Information Science Research Centers, Superconducting Quantum Materials
and Systems Center (SQMS) under contract No. DE-AC02-07CH11359. M.S.A. acknowledges support from USRA NASA Academic Mission Services under contract No. NNA16BD14C with NASA, with this work funded under the NASA-DOE
interagency agreement SAA2-403602 governing NASA’s work as part of the SQMS center. The authors wish to thank Jason Saied and Ryan LaRose for useful discussions on Floquet and error correcting codes in general, as well as Nikolas Breuckmann, Christophe Vuillot and Margarita Davydova for helpeful comments during the final stages of this work. M.S.A. would also like to thank Alex Townsend-Teague and  Julio Magdalena de la Fuente for useful discussions, especially involving the ZX calculus, and especially Xiaozhen Fu, for enlightening correspondence on calculating the distance of Floquet codes.

\bibliography{refs}

\appendix
\input{./appendix}

\end{document}

%% file: appendix.tex
\section{Preservation of logical information}

Here, we expand on the discussion on logical preservation between any two stabilizer codes. In particular, this discussion applies to the ISGs of the Floquet-Bacon-Shor family of codes described above. Let us suppose that $\ket{b^A}$ ($\ket{b^B}$) is a codeword such that it is an eigenvector of the logical operator $Z^A$ ($Z^B$) of code $A$ ($B$) with eigenvalue $\alpha$. The following short calculation shows that the projected codeword $\ket{b'^{B}} = \prod_{s \in \mathcal{S}^B} \left( \frac{\mathbb{I} + s}{2}\right)\ket{b^A}$ is an eigenvector of the logical operator $Z^B$ of code $B$ with the same eigenvalue $\alpha$ whenever $s^A Z^A = s^B Z^B$ for some $s^A \in \mathcal{S}^A$ and $s^B \in \mathcal{S}^B$.
\begin{eqnarray}
Z^B \ket{b'^{B}} &=& Z^B \prod_{s \in \mathcal{S}^B} \left( \frac{\mathbb{I} + s}{2}\right) \ket{b^A} \nonumber \\
&=&  \prod_{s \in \mathcal{S}^B} \left( \frac{\mathbb{I} + s}{2}\right)  s^B Z^B \ket{b^A} \nonumber \\
&=& \prod_{s \in \mathcal{S}^B} \left( \frac{\mathbb{I} + s}{2}\right)  s^A Z^A \ket{b^A} \nonumber \\
&=& \prod_{s \in \mathcal{S}^B} \left( \frac{\mathbb{I} + s}{2}\right)  \alpha \ket{b^A} \nonumber \\
&=& \alpha \ket{b'^{B}}
\end{eqnarray}
In other words, the projected codeword $\ket{b'^{B}}$ behaves exactly like a codeword of code $B$, i.e. $\ket{b'^{B}} = \ket{b^B}$. In turn, this implies that when we project the state $\ket{\psi^A} = \alpha \ket{0^{A}} + \beta \ket{1^A}$ to $\ket{\psi^B} = \alpha \ket{0'^{B}} + \beta \ket{1'^B}$, we will ensure that $Z^B \ket{0'^{B}} = \ket{0'^{B}}$ and $Z^B\ket{1'^B} = -\ket{1'^B}$, so that while the logical codewords $\ket{0^B} = \ket{0'^B}$ and $\ket{1^B} = \ket{1'^B}$ have deformed, the information has remained intact.

As a simple illustration of this, consider the two familiar 3-qubit repetition codes
\begin{eqnarray}
\mathcal{S}^{(0)} &=& \langle Z_1 Z_2, Z_2 Z_3 \rangle \nonumber \\
\mathcal{S}^{(1)} &=& \langle X_1 X_2, X_2 X_3 \rangle
\end{eqnarray}
with the same logical operators for both codes
\begin{eqnarray}
\overline{X} &=& X_1 X_2 X_3 \nonumber \\
\overline{Z} &=& Z_1 Z_2 Z_3
\end{eqnarray}
and codewords
\begin{eqnarray}
\ket{\overline{0}}^{(0)} = \ket{000} &,& \ket{\overline{0}}^{(1)} = \frac{1}{\sqrt{2}} \left( \ket{+++} + \ket{---} \right) \nonumber \\
\ket{\overline{1}}^{(0)} = \ket{111} &,& \ket{\overline{1}}^{(1)} = \frac{1}{\sqrt{2}} \left( \ket{+++} - \ket{---} \right)
\end{eqnarray}
It is clear that the logical operators defined above satisfy the requirements of preserving logical information from one round to the next, by simply picking $s^{(0)} = s^{(1)} = \mathbb{I}$. 
Note that this is why we require the unconventional definition of the logical operators for $\mathcal{S}^{(1)}$ which are usually defined the other way round, i.e. $\overline{X} \Leftrightarrow \overline{Z}$. Similarly, the codewords for $S^{(1)}$ are different from the conventional $\ket{\overline{0}} = \ket{+++}$ and $\ket{\overline{1}} = \ket{---}$.\\

If we view the two repetition codes as ISGs of a 3-qubit Floquet code, then even though logical information is conserved from one round to the next, the code distance is 1 since no stabilizers remain preserved between rounds and this cannot even detect single-qubit $Z$ ($X$) errors, so that such errors become logical errors that persist once they occur.

We also note that Eq.~\eqref{eqn:preserve-info} guarantees that the commutation structure of the logical operators remains intact. Using identities $[A, BC] = B[A,C] + [A,B]C$ and $[AB,C] = A[B,C] + [A,C]B$, we can quickly prove that $[O_{1}^{A}, O_{2}^{A}] = 0 \Rightarrow [s_{1}^{A}O_{1}^{A}, s_{2}^{A}O_{2}^{A}] = 0$, for any $s_{1}^{A}, s_{2}^{A} \in \mathcal{S}^{A}$. 
If we have $s_{1}^{A}O_{1}^{A} = s_{1}^{B}O_{1}^{B}$ and $s_{2}^{A}O_{2}^{A} = s_{2}^{B}O_{2}^{B}$, then under the assumption of Eq.~\eqref{eqn:preserve-info}, this implies that
\begin{equation}
    [s_{1}^{B} O_{1}^{B}, s_{2}^{B}O_{2}^{B}] = 0
\end{equation}
Again, using commutator identities, and the facts that $O_{1,2}^{B}$ are logical operators at round $B$, i.e. $[O_{1,2}^{B}, s^{B}] = 0$ for all $s^{B} \in \mathcal{S}^{B}$, we find that this implies $[O_{1}^{B}, O_{2}^{B}] = 0$. The case for the anti-commutator is proved similarly, using anti-commutator identities $\{A,BC\} = [A,B]C + B\{A,C\}$ and $\{AB, C \} = A\{B,C \} - [A,C]B = A[B,C] + \{A,C\}B$.

\section{Self-correction}
Suppose that $X$ errors occur on two qubits that lie on the row above cell A on either side of the plaquette column AD between rounds 0 and 1 mod 4. Such a 2-qubit error is equivalent, upto stabilizer transformations of the ISG at round 0 mod 4 to a 2-qubit $XX$ error that occurs on the top edge of cell A. This anti-commutes with the logical $\overline{Z}$ error of the dynamical qubit at round 0 mod 4, and thus contains a factor of the logical $\overline{X}$ operator. Specifically, it is given by $\overline{X}_A \overline{X}_{A'}$, where cell $A'$ is defined as the cell on top of cell A. In the next round 1 mod 4, we would measure a subset of all the $ZZ$ checks which, in addition to other transient stabilizers, would measure the gauge qubit operator $\overline{Z}_{A'}$, as well as the check operator $\overline{Z}_A \overline{Z}_B$. Suppose, without loss of generality, that in the absence of any errors, the measurements of both would have yielded result +1. Due to the error, their values will flip. In the next round 2 mod 4, we measure the 2-qubit check $\overline{X}_A \overline{X}_{A'}$, and use the property that $(\mathbb{I} \pm \sigma) = \pm(\mathbb{I} \pm \sigma) \sigma$ for any Pauli operator $\sigma$. In the measurement round 2 mod 4, this check takes on some random value $\pm 1$. The entire sequence of steps can be followed as
\begin{widetext}
\begin{eqnarray}
\ket{\psi^{(0)}} &\xrightarrow{\text{error}}& \overline{X}_A \overline{X}_{A'}\ket{\psi^{(0)}} \nonumber \\
&\xrightarrow{\text{detection}}& \left[ \dots (\mathbb{I} - \overline{Z}_{A} \overline{Z}_{B})(\mathbb{I} - \overline{Z}_{A'})\right] \overline{X}_A \overline{X}_{A'}\ket{\psi^{(0)}} \nonumber \\
&\xrightarrow{\text{measurement}}& \left[ \dots (\mathbb{I} \pm \overline{X}_A \overline{X}_{A'})\right]  \left[ \dots (\mathbb{I} - \overline{Z}_{A} \overline{Z}_{B})(\mathbb{I} - \overline{Z}_{A'})\right] \overline{X}_A \overline{X}_{A'}\ket{\psi^{(0)}} \nonumber \\
&=& \pm \left[ \dots (\mathbb{I} \pm \overline{X}_A \overline{X}_{A'} )\right] \overline{X}_A \overline{X}_{A'} \left[ \dots (\mathbb{I} - \overline{Z}_{A} \overline{Z}_{B})(\mathbb{I} - \overline{Z}_{A'})\right] \overline{X}_A \overline{X}_{A'}\ket{\psi^{(0)}} \nonumber \\
&=& \pm \left[ \dots (\mathbb{I} \pm \overline{X}_A \overline{X}_{A'} )\right] \left[ \dots (\mathbb{I} + \overline{Z}_{A} \overline{Z}_{B})(\mathbb{I} + \overline{Z}_{A'})\right] \ket{\psi^{(0)}} \nonumber \\
&=& \pm \left[ \dots (\mathbb{I} \pm \overline{X}_A \overline{X}_{A'} )\right] \ket{\psi^{(1)}} \nonumber \\
&=& \pm\ket{\psi^{(2)}}
\end{eqnarray}
\end{widetext}
so that the entire situation is indistinguishable from reaching the correct logical state $\ket{\psi^{(2)}}$ at round 2 mod 4 without any errors.

The above discussion also implies that the correcting operator for a single $X$ error on any single row is the same whether it lies on the left or the right of the blue domain wall, even though they produce distinct syndromes. Suppose that such a single qubit $X$ error occurred on the right side of the blue domain wall along column AD on the row above cell A between rounds 0 and 1 mod 4. Let B' be the cell above B, and $s_{AB}$ and $s_{A'B'}$ denote the permanent $Z$-type stabilizers along plaquette rows $AB$ and $A'B'$ respectively. Then, this error is equivalent to $\overline{X}_{s_{AB}} \overline{X}_{s_{A'B'}} \overline{X}_{A} \overline{X}_{A'}$, up to stabilizer transformations of the ISG at round 0 mod 4, where the product $\overline{X}_{s_{AB}} \overline{X}_{s_{A'B'}}$ gives a single qubit $X$ operator that lies on the left-most column on this row. At round 1 mod 4, this error will flip the value of the permanent $Z$-type stabilizer $\overline{Z}_{s_{A'B'}}$ as well as the transient stabilizer $\overline{Z}_{A'}$. Note that the permanent $Z$-type stabilizer $\overline{Z}_{s_{AB}}$ does not get measured in this round. Suppose, without loss of generality, that in the absence of such an error, both $\overline{Z}_{s_{A'B'}}$ and $\overline{Z}_{A'}$ would have measured out to be +1, so that this error flips the recorded values to be -1 instead.

A correction operation of $\overline{X}_{s_{AB}} \overline{X}_{s_{A'B'}}$ suffices after round 1 mod 4, as the subsequent measurements of round 2 mod 4 then restore the correct logical state, as can be seen through the following sequence of transformations
\begin{widetext}
\begin{eqnarray}
\ket{\psi^{(0)}} &\xrightarrow{\text{error}}& \overline{X}_{s_{AB}} \overline{X}_{s_{A'B'}} \overline{X}_A \overline{X}_{A'}\ket{\psi^{(0)}} \nonumber \\
&\xrightarrow{\text{detection}}& \left[ \dots (\mathbb{I} - \overline{Z}_{s_{A'B'}}) (\mathbb{I} - \overline{Z}_{A'}) \right] \overline{X}_{s_{AB}} \overline{X}_{s_{A'B'}} \overline{X}_A \overline{X}_{A'}\ket{\psi^{(0)}} \nonumber \\
&\xrightarrow{\text{correction}}& \overline{X}_{s_{AB}} \overline{X}_{s_{A'B'}} \left[ \dots (\mathbb{I} - \overline{Z}_{s_{A'B'}}) (\mathbb{I} - \overline{Z}_{A'}) \right] \overline{X}_{s_{AB}} \overline{X}_{s_{A'B'}} \overline{X}_A \overline{X}_{A'}\ket{\psi^{(0)}} \nonumber \\
&\xrightarrow{\text{measurements}}& \left[ \dots (\mathbb{I} \pm \overline{X}_A \overline{X}_{A'})\right] \overline{X}_{s_{AB}} \overline{X}_{s_{A'B'}} \left[ \dots (\mathbb{I} - \overline{Z}_{s_{A'B'}}) (\mathbb{I} - \overline{Z}_{A'}) \right] \overline{X}_{s_{AB}} \overline{X}_{s_{A'B'}} \overline{X}_A \overline{X}_{A'}\ket{\psi^{(0)}} \nonumber \\
&=& \pm \left[ \dots (\mathbb{I} \pm \overline{X}_A \overline{X}_{A'})\right] \overline{X}_A \overline{X}_{A'} \overline{X}_{s_{AB}} \overline{X}_{s_{A'B'}} \left[ \dots (\mathbb{I} - \overline{Z}_{s_{A'B'}}) (\mathbb{I} - \overline{Z}_{A'}) \right] \overline{X}_{s_{AB}} \overline{X}_{s_{A'B'}} \overline{X}_A \overline{X}_{A'}\ket{\psi^{(0)}} \nonumber \\
&=& \pm \left[ \dots (\mathbb{I} \pm \overline{X}_A \overline{X}_{A'})\right] \left[ \dots (\mathbb{I} + \overline{Z}_{s_{A'B'}}) (\mathbb{I} + \overline{Z}_{A'}) \right] \ket{\psi^{(0)}} \nonumber \\
&=& \pm \left[ \dots (\mathbb{I} \pm \overline{X}_A \overline{X}_{A'})\right] \ket{\psi^{(1)}} \nonumber \\
&=& \pm \ket{\psi^{(2)}} \nonumber \\
\end{eqnarray}
\end{widetext}
Note that the same correction operation would also work if the factor of $\overline{X}_A \overline{X}_{A'}$ was absent from the error, and it had instead occurred on the left side of the blue domain wall on that row, which would have been equivalent to $\overline{X}_{s_{AB}} \overline{X}_{s_{A'B'}}$ upto transformations of the ISG at round 0 mod 4.

\section{Unmasked Code distance}
\label{appdx:code-distance}
Here, we describe in detail the application of the distance algorithm of \cite{fu2024errorcorrectiondynamicalcodes} to our construction of Floquet-Bacon-Shor codes. In particular, we compute the unmasked distance, defined as the centralizer of certain unmasked stabilizers, excluding a suitably identified gauge group, both of which are identified with the aid of the distance algorithm. The unmasked stabilizers could be thought of as the detectors that can form from the set of elements of a given ISG, while the gauge group roughly corresponds to the set of measurements in the current round and their destabilizers from the next round, both of which would commute with any such detectors, and therefore belong in the centralizer of the unmasked stabilizers, but do not constitute logical errors.

The distance algorithm works by sequentially updating two sets, labelled $V$ and $C$. In addition, we also initialize two empty sets $\tilde{P}$ and $\tilde{U}$. The set $V$ is initialized as the trivial stabilizer group $\langle I \rangle$, and updated according to the rules of ISG growth. In particular, if a Pauli operator $m$ is measured, then
\begin{itemize}
\item (A) If $[m,v] =0$ $\forall v \in V$, then $V$ is updated to $V' = \langle V, m \rangle$. Note that $V$ remains unchanged if $m \in V$ already, but increases its rank by 1 if $m \not\in V$.
\item (B) If $\{m,v_j\}=0$ for some subset ${v_j} \subset V$, then the first of such elements (in any arbitrary ordering) $v_1$ is removed from $V$ and the rest updated according to $v_j \rightarrow v_j v_1$, so that $[m, v_j v_1] =0$. We also add $m$ to $V$, so that in all, the rank of $V$ remains unchanged in this case.
\end{itemize}

Meanwhile, the set $C$ is initialized to some given ISG, with respect to which the unmasked distance is computed, and its elements are only ever removed over the course of the algorithm, never added. In particular,
\begin{itemize}
\item (C) If $[m,c]=0$ $\forall c \in C$, then $C$ is left unchanged.
\item (D) If $\{m,c\}=0$ for some $c \in C$, and
    \begin{itemize}
    \item (i) $[m,v]=0$ $\forall v \in V$, then we update all $c_i \neq c$ such that $\{m, c_i\} = 0$ as $c_i \rightarrow c \cdot c_i$, remove $c$ from $C$ and add $c$ to the (initially empty) set $\tilde{P}$
    \item (ii) $\{m,v_j\}=0$ for some subset $\{v_j\}_{j=1}^{|J|} \subset V$, then having removed $v_1$ from $V$ by rule (B) above, we update all $c_j \in C$ that anti-commute with $m$ as $c_j \rightarrow v_1 \cdot c_j$ (note that this somewhat similar to how a logical operator is updated).
    \end{itemize}
\end{itemize}

Finally, at any stage of the measurement schedule, we update the (initially empty) set $\tilde{U}$ that contains two types of elements
\begin{itemize}
\item All elements in $\langle C \rangle \cap \langle V \rangle$,
\item All elements $c \in C$ that get updated to the identity, i.e. $c \rightarrow c \cdot v = \mathbb{I}$ according to rule (D)(ii) above.
\end{itemize}

Roughly speaking, these correspond to the elements in the original ISG, which was identified with $C$ at the very beginning of the algorithm, that are able to form detectors. These detectors are formed by comparison of the measured values of the same operator in two different rounds, which allows us to readily identify $\tilde{U}=U$, the set of all unmasked stabilizers. Algorithm 2 of \cite{fu2024errorcorrectiondynamicalcodes} also allows for a more general case where detectors can form from measurements across several rounds, and we refer the reader to that paper for further details.

Similarly, the set $\tilde{P}$ contains elements, which along with suitably identified destabilizers according to Alogrithm 3 of \cite{fu2024errorcorrectiondynamicalcodes} allows us to identify the gauge group, which in our case is also simplified in our case due to working with the virtual qubit operators directly.


\begin{figure*}[!tp]
    \centering
\includegraphics[width=\textwidth]{distance_algorithm.jpg}
\caption{The distance algorithm of \cite{fu2024errorcorrectiondynamicalcodes} applied to a Floquet-Bacon-Shor code with a single dynamical logical qubit. Unlike in previous figures, here we depict the permanent Bacon-Shor stabilizers in addition to the virtual operators of the gauge qubits for convenience.}
\label{fig:distance-algorithm}
\end{figure*}

We apply this algorithm to an example Floquet-Bacon-Shor code on a $7 \times 7$ lattice, computing the unmasked distance of the ISG $S_{4k}$ for $k \in \mathbb{Z}$, as depicted in Fig. \ref{fig:distance-algorithm}. In the notation there, the empty grid represents the trivial group $\langle I \rangle$, while colored grids represent a group generated by the denoted virtual qubit operators. The set $V_0$ is initialized to the empty grid, while the set $C_0$ is identified with the ISG $S_{4k}$.

At each of the rounds, we measure all the virtual qubit operators as depicted, including those for the stabilizer qubits of the Bacon-Shor subsystem code. At round 1, this updates the set $V_1$ to exactly equal the measured set $M_1$ using rule (A) above, while elements from $C$ are removed using rule (D)(i) and dumped into the set $\tilde{P}_1$, which remains unchanged (i.e. $\tilde{P}_j = \tilde{P}_1$ $\forall j$) after this step. The set $\tilde{U}_1$ contains the intersection of $C_1$ and $V_1$ at this stage. 

At round 2, the measured operators update the set $V_2$ to almost coincide with the steady state ISG $S_{4k+2}$, except we have not yet measured one $X$-type and one $Z$-type permanent Bacon-Shor stabilizers yet. Thet set $C_2$ has all the $Z$ (blue) gauge operators along the defect column updated to the identity using rule (D)(ii), and therefore these now persist in $\tilde{U}_2$, which in addition also now contains $X$ (red) gauge operators along a defect row since they belong in the intersection of $C_2$ and $V_2$.

At round 3, these $X$ (red) gauge operators are updated to the identity using rule (D)(ii) and therefore persist in $\tilde{U}_3$, which now also includes all the permanent $Z$ (blue) Bacon-Shor stabilizers, since these belong in the intersection of $C_3$ and $V_3$. The latter is now almost identical with the steady state ISG $S_{4k+3}$, except it does not contain one of the $X$ (red) permanent Bacon-Shor stabilizers, while the set $C_3$ now only contains all the permanent Bacon-Shor stabilizers. Any number of measurements beyond this will not change the structure of $C_j$, as it will always contain these, and only these, permanent Bacon-Shor stabilizers for all $j \geq 3$.

Finally, at round 4, completing one measurement cycle, the set $V_4$ coincides exactly with the steady state ISG $S_{4(k+1)}$, the previously noted $X$ gauge operators along the defect row in $\tilde{U}_3$ persist in $\tilde{U}_4$, since they now belong in the intersection of $V_4$ and $C_4$. The colored generators shown in $\tilde{U}_4$ precisely generate the group of unmasked stabilizers $U$.

Meanwhile, the gauge group is generated by the generators shown in $\tilde{P}$ along with a suitable choice of destabilizers, in addition to the unmasked stabilizers themselves. These are depicted in the bottom part of Fig. \ref{fig:distance-algorithm}. The unmasked distance is the minimum weight of any operator in the centralizer of the unmasked stabilizers that excludes all elements in this gauge group (including the unmasked stabilizers themselves). The unmasked stabilizer are given by the colored boxes in $\tilde{U}_4$ in Fig. \ref{fig:distance-algorithm}. Clearly, any operator corresponding to the same boxes with the opposite color cannot be part of the centralizer, so we exclude such operators. All colored boxes included in the definition of $\mathcal{G}$ as shown in Fig. \ref{fig:distance-algorithm} are also excluded by definition. This leaves a single box, the top box in the gauge defect (which is always composed of two boxes in any round and features a single check, a red $XX$ check in this case, between those two boxes). Either an $\overline{X}$ or $\overline{Z}$ operator for this box, or gauge qubit, is an operator that normalizes the unmasked stabilizer, and is not in the gauge group. The minimum weight of such an operator is $d-1$ on a $d \times d$ square lattice (that of the $\overline{X}$ in this example), which is therefore the unmasked distance.


\section{Numerical simulations}
\label{appdx:numerics}
Here, we describe the numerical simulations carried out to produce Figs. \ref{fig:code-capacity}, \ref{fig:code-capacity-distance}, \ref{fig:kdn-scaling} and \ref{fig:fbs-repeated-rounds} in more detail. First, we describe the simulation performed for the Bacon-Shor code in the code capacity setting in pseudo-code form in Algorithm \ref{algo:bs-code-capacity}. The approach used for simulating the Floquet-Bacon-Shor under the same noise model is similarly described in pseudo-code in Algorithm \ref{algo:fbs-code-capacity}.

For the Bacon-Shor code, we initialize all qubits in the $\ket{0}$ state, thereby fixing all physical $Z$ operators to $+1$ while also fixing the value of the logical $Z$ operator to be $+1$. After all $X$-type checks are measured in the first round 0, only products of the individual $Z$ operators survive, specifically the products that form the $Z$-type stabilizers of the Bacon-Shor code. At the first round 1, these stabilizers are re-measured and can therefore form detectors as simply the product of all the measured checks in this round forming a stabilizer, without requiring comparison to previous measurements since all qubits were initialized in the $\ket{0}$ state. After this initial round, we repeat the pattern of measuring $X$ and then $Z$ checks for several measurement cycles, each time forming detectors from $X$ and $Z$ stabilizers respectively. Finally, we read out all qubits in the computational basis, once again measuring all physical $Z$ operators as we did when we initialized the qubits in the computational basis, but with non-deterministic outcomes this time. However, the $Z$ checks measured in the very last round 1, prior to these single qubit readouts, do survive upon these measurements and can therefore form detectors. Without these last set of checks as detectors, a single $X$ error after the last round 1 measurements and before the final qubit readouts would cause an undetectable logical error, and this last round of checks as detectors are therefore important in retaining the code distance. To simulate noise, single qubit depolarizing channels are applied to all qubits before each ``Measure all ($X$ or $Z$) checks of round (0 or 1)" step or the final ``Qubit readouts" step.

\begin{algorithm}
\SetKwBlock{Initial}{Initial measurement cycle}{end}  
\SetKwBlock{MeasureRoundZero}{Measure all $X$-checks of round 0}{end}  
\SetKwBlock{MeasureRoundOne}{Measure all $Z$-checks of round 1}{end}  
\SetKwBlock{Repeat}{Repeat for $M$ measurement cycles}{end}  
\SetKwBlock{Final}{Qubit readouts}{end}  
\Initial{
Initialize all qubits in the $\ket{0}$ state\;
1q depolarizing noise\;
\MeasureRoundZero{(Do nothing else in this initial round)\;}{}
1q depolarizing noise\;
\MeasureRoundOne{
    Form detectors out of all $Z$-type stabilizers (without comparison to prior measurements, since qubits were reset in the $Z$ basis)\;}{}
}

\Repeat{
    1q depolarizing noise\;
    \MeasureRoundZero{Form detectors out of all $X$-type detectors\;}{}
    1q depolarizing noise\;
    \MeasureRoundOne{Form detectors out of all $Z$-type detectors\;}{}
}{}

\Final{
    1q depolarizing noise\;
    Measure all qubits in the computational ($Z$) basis\;
    Form detectors out of measured checks from round 1 of the last measurement cycle, comparing to the current individual qubit measurements\;
    Measure the logical $Z$ operator as the product of single qubit measurements along any row (e.g. the bottom most row of the lattice)\;
}{}
\caption{Bacon-Shor: code capacity}
\label{algo:bs-code-capacity}
\end{algorithm}

We proceed fairly similarly for the Floquet-Bacon-Shor code. We initialize all qubits in the $\ket{0}$ state, thereby fixing the value of both the dynamical logical $Z$ operator for the single dynamical qubit in this experiment, as well as the static logical $Z$ operator inherited from the parent Bacon-Shor subsystem code to $+1$. After this initialization, we measure all $X$ checks of round 0, with which the temporary as well as permanent stabilizers of round 1, as described in the main text, commute. These serve as detectors in the first iteration of round 1, without the need for comparison to prior measurements, since these products were already fixed upon the initialization of all qubits. After this, we measure all $X$ checks of round 2, but can only form detectors out of the temporary stabilizers described in the main text, comparing to their previously measured values at round 0. We cannot construct a detector out of the permanent stabilizer measured in this round along column AD since there is no prior measurement to compare it against. In round 3 of the initial measurement cycle, we can form detectors out of the temporary stabilizers by comparing to their prior measurement at round 1 of this initial cycle, but we can also form a detector out of the permanent stabilizer measured in this round, without the need for comparing to prior measurements, since this was also implicitly measured when we initialized the qubits to the $\ket{0}$ state.

After this initial cycle, we repeatedly measure all checks from the four rounds in sequence. Before each round, we apply single qubit depolarizing noise channels on all qubits. Using the measurements at each round, we form detectors out of both temporary and permanent stabilizers measured in that round, and update the dynamical logical $Z$ operator by multiplying it with appropriate measurements as given in Table \ref{table:logical-ops}. Finally, we read out all the qubits in the computational basis, once again measuring all physical $Z$ operators, and therefore also the values of the dynamical logical $Z$ operator, given as the gauge operator $Z_C$ since this specifies the instantaneous logical operator at round 3, the last measured round before the qubit readouts, as well as the static logical $Z$ of the parent Bacon-Shor subsystem code. We also form detectors out of all checks measured at the last round 3, and also form a detector out of the permanent stabilizer along row CD using the individual qubit readout measurements as well as the producrt of measurements that specified this stabilizer from the last round 1. As before, these last set of checks help identify any errors resulting from the very last depolarizing round before the individual qubit readouts. Together, the results of these two simulations produce Fig. \ref{fig:code-capacity}.

We use a very similar setup as Algorithm \ref{algo:fbs-code-capacity} to compute the scaling of $kd/n$ in Fig. \ref{fig:code-capacity-distance} for the Floquet-Bacon-Shor code. The major difference is that we now have $k = \Theta(n)$ many logical operators to readout at the end, and roughly as many detectors to form as well. As before, these detectors are formed out of both the temporary and stabilizer stabilizers. In this case, where we have multiple dynamical qubits, there are as many permanent stabilizer detectors in each round as there are defect columns or rows, and all the gauge fixed operators along those defect columns or rows are the temporary stabilizers that we form detectors out of. The pattern of initial and final detectors out of either the initialization of the qubits, or their final readouts (both fixing the physical $Z$ operators) is the same as before, except we form many more detectors in this case. 

For our simulation, we kept $d$ constant while increasing $k = q^2 = \Theta(n)$ as a function of the parameter $q$, though other combinations are also possible. Targeting $d=4$ as described in the main text translates to having a single square plaquette between the lattice boundary and the closest defect or between neighboring defects. We have $q$ many defect columns or rows in each round, each of which host $q+1$ temporary stabilizers, since each defect is one square plaquette away from each other in that same defect column/row. We can form $q+1$ detectors out of these $q$ many temporary stabilizers, by forming products of the checks between pairs of neighboring defects, or a boundary defect and the lattice boundary. Together, these give $(q+1)^2$ many detectors. In addition, we also form detectors out of the $q$ many permanent stabilizers measured at each round, giving a total of $(q+1)^2 + q$ many detectors at each round during the repeat cycle. We used Stim's $\texttt{shortest\_graphlike\_error}$ function to compute the effective code distance of each code in the family parameterized by $q$, and found it to be exactly $4$ in all cases. The ratio $kd/n$ was then computed using $k = q^2 + 1$, $d=4$ and $n = (3q+2)^2$, and plotted in Fig. \ref{fig:code-capacity-distance}.

Finally, in order to produce Fig. \ref{fig:fbs-repeated-rounds} we performed two simulations. One carried out the same sequence of operations as in Algorithm \ref{algo:fbs-code-capacity} except that we introduce reset errors as bit-flip channels on individual qubits after resetting all qubits to the $\ket{0}$ state, have measurement errors for each measured check, and also have measurement errors for the individual qubit readouts at the very end. The other simulation was carried out usual the ``repeated rounds" schedule, in which we carry out only a single measurement cycle in the ``Repeat" phase, each measurement round is repeated $4d$ many times, where $d$ is the grid diameter. The usual detectors formed as in Algorithm \ref{algo:fbs-code-capacity} only in the first of such $4d$ many rounds. In the remaining rounds, we instead form detectors out of the individual measured check operators. This increases the density of detectors and gives better logical error rates than if we had continued to form detectors out of the usual temporary and permanent stabilizers instead. Similar ideas were explored in \cite{higgott2021subsystem} to improve performance in subsystem codes. To allow for a steady state ISG sequence, we allow for an initial measurement cycle with no repeated rounds before the ``Repeat" phase, and to allow simpler syndrome exctraction during the qubit readouts, we allow for another measurement cycle without any repeated rounds after the ``Repeat" cycle. In all, we have $4(d+2)$ many measurement rounds in the ``repeated rounds" cycle, $4d$ of which occur during the ``Repeat" phase in a single measurement cycle. Meanwhile, we have $4d$ measurement rounds, divided across $4d$ measurement cycles in the usual schedule. In both cases, we divide the total number of sampled logical errors by the product of the number of shots and the total number of measurement rounds and plot this as the logical error rate per measurement round in Fig. \ref{fig:fbs-repeated-rounds}.

\begin{algorithm}
\SetKwBlock{Initial}{Initial measurement cycle}{end}  
\SetKwBlock{MeasureRoundZero}{Measure all $X$-checks of round 0}{end}  
\SetKwBlock{MeasureRoundOne}{Measure all $Z$-checks of round 1}{end}  
\SetKwBlock{MeasureRoundTwo}{Measure all $X$-checks of round 2}{end}  
\SetKwBlock{MeasureRoundThree}{Measure all $Z$-checks of round 3}{end}  
\SetKwBlock{Repeat}{Repeat for $M$ measurement cycles}{end}  
\SetKwBlock{Final}{Qubit readouts}{end}  

\Initial{
Initialize all qubits in the $\ket{0}$ state\;
1q depolarizing noise\;
\MeasureRoundZero{(Do nothing else in this initial round)\;}{}
1q depolarizing noise\;
\MeasureRoundOne{
    Form detectors out of all temporary and permanent stabilizers (without comparison to prior measurements, since qubits were reset in the $Z$ basis)\;
    Update dynamical logical $Z$-operator\;}{}
1q depolarizing noise\;
\MeasureRoundTwo{
    Form detectors out of all temporary stabilizers, comparing to their previously measured value at round 0\;
    Update dynamical logical $Z$-operator\;}{}
1q depolarizing noise\;
\MeasureRoundThree{
    Form detectors out of all temporary and permanent stabilizers (without comparison to prior measurements, since qubits were reset in the $Z$ basis)\;
    Update dynamical logical $Z$-operator\;}{}
}{}

\Repeat{
    1q depolarizing noise\;
    \MeasureRoundZero{Form all detectors\;Update dynamical logical $Z$-operator;}{}
    1q depolarizing noise\;
    \MeasureRoundOne{Form all detectors\;Update dynamical logical $Z$-operator;}{}
    1q depolarizing noise\;
    \MeasureRoundTwo{Form all detectors\;Update dynamical logical $Z$-operator;}{}
    1q depolarizing noise\;
    \MeasureRoundThree{Form all detectors\;Update dynamical logical $Z$-operator;}{}
}{}

\Final{
    1q depolarizing noise\;
    Measure all qubits in the computational ($Z$) basis\;
    Form detectors out of measured checks from round 3 of the last measurement cycle, comparing to the current individual qubit measurements\;
    Form detectors out of the last measured value of the permanent stabilizer along (plaquette) row CD from round 1 of the last measurement cycle, comparing to the current individual qubit measurements\;
    Measure the dynamical logical $Z$ operator as the gauge operator $\overline{Z}_C$ using the measured values of individual qubits\;
    Measure the static logical $Z$ operator as the product of single qubit measurements along any row (e.g. the bottom most row of the lattice)\;
}{}
\caption{Floquet-Bacon-Shor: code capacity}
\label{algo:fbs-code-capacity}
\end{algorithm}